\documentclass[sigconf, nonacm]{acmart}

\newcommand\vldbdoi{10.14778/3446095.3446096}
\newcommand\vldbpages{XXX-XXX}
\newcommand\vldbvolume{14}
\newcommand\vldbissue{5}
\newcommand\vldbyear{2021}
\newcommand\vldbauthors{\authors}
\newcommand\vldbtitle{\shorttitle} 
\newcommand\vldbavailabilityurl{}
\newcommand\vldbpagestyle{plain}

\def\fullversion{}

\usepackage{amsmath,amsfonts}
\usepackage{graphicx}
\usepackage{epstopdf}
\usepackage{wrapfig}
\usepackage{color,colortbl}
\usepackage{xcolor}
\usepackage{url}
\usepackage{subfigure}
\usepackage{xspace}
\usepackage{balance}
\usepackage{times}
\usepackage{comment}

\usepackage{float} 
\usepackage{hyperref}

\usepackage[noend]{algorithmic}
\usepackage[boxed]{algorithm}

\newtheorem{lemma}{Lemma}
\newtheorem{theorem}[lemma]{Theorem}
\newtheorem{definition}[lemma]{Definition}
\newtheorem{corollary}[lemma]{Corollary}
\newtheorem{proposition}[lemma]{Proposition}

\newcommand{\stitle}[1]{\noindent{\bf #1}}

\renewcommand{\qed}{\hfill $\framebox(6,6){}$}
\newcommand{\eop}{{\hspace*{\fill}$\Box$\par}}

\newcommand{\ctab}{Table~}

\newcommand{\csec}{Section~}

\newcommand{\cthm}{Theorem~}

\newcommand{\cprop}{Proposition~}

\newcommand{\ccor}{Corollary~}

\newcommand{\cfig}{Figure~}
\newcommand{\cfigs}{Figures~}
\newcommand{\capp}{Appendix~}

\newcommand{\ie}{{\em i.e.}\xspace}
\newcommand{\eg}{{\em e.g.}\xspace}

\newcommand{\etal}{{\em et al.}\xspace}

\newcommand{\squishlist}{
	\begin{list}{$\bullet$}{
		\setlength{\itemsep}{0pt}
		\setlength{\parsep}{3pt}
		\setlength{\topsep}{3pt}
		\setlength{\partopsep}{0pt}
		\setlength{\leftmargin}{1.0em}
		\setlength{\labelwidth}{1em}
		\setlength{\labelsep}{0.5em}
   }
}

\newcommand{\squishenum}{
	
	\begin{list}{\usecounter{scount}}{
		\setlength{\itemsep}{0pt}
		\setlength{\parsep}{3pt}
		\setlength{\topsep}{3pt}
		\setlength{\partopsep}{0pt}
		\setlength{\leftmargin}{1.2em}
		\setlength{\labelwidth}{1em}
		\setlength{\labelsep}{0.5em}
	}
}

\newcommand{\squishend}{
	\end{list}
}

\newcommand{\eat}[1]{}

\newcommand{\pr}[1]{{\bf Pr}\left[#1\right]\xspace}
\newcommand{\prinline}[1]{{\bf Pr}[#1]\xspace}
\newcommand{\ep}[2][]{{\bf E}_{#1}\hspace{-0.06cm}\left[#2\right]\xspace}
\newcommand{\epinline}[1]{{\bf E}[#1]\xspace}

\newcommand{\vr}[1]{{\bf Var}\hspace{-0.06cm}\left[#1\right]\xspace}
\newcommand{\vrinline}[1]{{\bf Var}[#1]\xspace}

\newcommand{\cvinline}[1]{{\bf Cov}[#1]\xspace}

\newcommand{\bigohinline}[1]{{\rm O}(#1)\xspace}

\newcommand{\bigthetainline}[1]{{\rm \Theta}(#1)\xspace}

\newcommand{\supind}[1]{^{(#1)}}

\newcommand{\eps}{\epsilon\xspace}

\definecolor{ao(english)}{rgb}{0.0, 0.5, 0.0}
\definecolor{lightgreen(backgroud)}{RGB}{100,255,100}
\definecolor{sensitive(backgroud)}{RGB}{255,160,160}
\definecolor{sensitive}{RGB}{255,80,80}
\definecolor{revision}{RGB}{0,0,255}

\newcommand{\code}[1]{\textcolor{blue}{\tt #1}\xspace}
\newcommand{\pred}{{\sf C}\xspace}

\newcommand{\sysname}{{\sf {F\lowercase{lash}P}}\xspace}
\newcommand{\samplename}{{\rm GSW}\xspace}

\newcommand{\dima}{{\bf a}\xspace}
\newcommand{\meaa}{{\bf m}\xspace}
\newcommand{\dimv}{{a}\xspace}
\newcommand{\meav}{{m}\xspace}
\newcommand{\lowerb}{\underline{\theta}\xspace}
\newcommand{\upperb}{\overline{\theta}\xspace}

\begin{document}


\title{FlashP: An Analytical Pipeline for Real-time Forecasting of Time-Series Relational Data}



%
%
%
%


\author{Shuyuan Yan}
\affiliation{
  \institution{Alibaba Group}
}
\email{raul.ysy@alibaba-inc.com}
\author{Bolin Ding}
\authornote{Bolin Ding is the corresponding author.}
\affiliation{
  \institution{Alibaba Group}
}
\email{bolin.ding@alibaba-inc.com}
\author{Wei Guo}
\affiliation{
  \institution{Alibaba Group}
}
\email{lengchuan.gw@alibaba-inc.com}
\author{Jingren Zhou}
\affiliation{
  \institution{Alibaba Group}
}
\email{jingren.zhou@alibaba-inc.com}
\author{Zhewei Wei}
\affiliation{
  \institution{Renmin University of China}
}
\email{zhewei@ruc.edu.cn}
\author{Xiaowei Jiang}
\author{Sheng Xu}
\affiliation{
  \institution{Alibaba Group}
}
\email{xiaowei.jxw@alibaba-inc.com}
\email{xusheng.xs@alibaba-inc.com}

\begin{abstract}
Interactive response time is important in analytical pipelines for users to explore a sufficient number of possibilities and make informed business decisions. We consider a forecasting pipeline with large volumes of high-dimensional time series data. Real-time forecasting can be conducted in two steps. First, we specify the part of data to be focused on and the measure to be predicted by slicing, dicing, and aggregating the data. Second, a forecasting model is trained on the aggregated results to predict the trend of the specified measure.
While there are a number of forecasting models available, the first step is the performance bottleneck. A natural idea is to utilize sampling to obtain approximate aggregations in real time as the input to train the forecasting model. Our scalable real-time forecasting system \sysname (Flash Prediction) is built based on this idea, with two major challenges to be resolved in this paper: first, we need to figure out how approximate aggregations affect the fitting of forecasting models, and forecasting results; and second, accordingly, what sampling algorithms we should use to obtain these approximate aggregations and how large the samples are. 
We introduce a new sampling scheme, called \samplename sampling, and analyze error bounds for estimating aggregations using \samplename samples. We introduce how to construct compact \samplename samples with the existence of multiple measures to be analyzed.
We conduct experiments to evaluate our solution and compare it with alternatives on real data. 
\end{abstract}

\maketitle

\pagestyle{\vldbpagestyle}
\begingroup\small\noindent\raggedright\textbf{PVLDB Reference Format:}\\
\vldbauthors. \vldbtitle. PVLDB, \vldbvolume(\vldbissue): \vldbpages, \vldbyear.\\
\href{https://doi.org/\vldbdoi}{doi:\vldbdoi}
\endgroup
\begingroup
\renewcommand\thefootnote{}\footnote{\noindent
This work is licensed under the Creative Commons BY-NC-ND 4.0 International License. Visit \url{https://creativecommons.org/licenses/by-nc-nd/4.0/} to view a copy of this license. For any use beyond those covered by this license, obtain permission by emailing \href{mailto:info@vldb.org}{info@vldb.org}. Copyright is held by the owner/author(s). Publication rights licensed to the VLDB Endowment. \\
\raggedright Proceedings of the VLDB Endowment, Vol. \vldbvolume, No. \vldbissue\ %
ISSN 2150-8097. \\
\href{https://doi.org/\vldbdoi}{doi:\vldbdoi} \\
}\addtocounter{footnote}{-1}\endgroup

\ifdefempty{\vldbavailabilityurl}{}{
\vspace{.3cm}
\begingroup\small\noindent\raggedright\textbf{PVLDB Artifact Availability:}\\
The source code, data, and/or other artifacts have been made available at \url{\vldbavailabilityurl}.
\endgroup
}


\eat{
\begin{figure}[t]
\small
\begin{tabular}{|c|c|c||c|c||c|}
\hline
\cellcolor{Apricot}{\sf Age} & \cellcolor{Apricot}{\sf Gender} & \cellcolor{Apricot}{\sf Location} & \cellcolor{YellowGreen}{\sf NumClick} & \cellcolor{YellowGreen}{\sf Impression} & \cellcolor{SkyBlue}{\sf TimeStamp}
\\
$\dima\supind{1}$ & $\dima\supind{2}$ & $\dima\supind{3}$ & $\meaa\supind{1}$ & $\meaa\supind{2}$ & $t$
\\ \hline \hline
\cellcolor{Yellow} 30 & \cellcolor{Yellow} F & WA & \cellcolor{Yellow} 5 & 1.6min & \cellcolor{Yellow} 20200301
\\ \hline
60 & M & WA & 1 & 1.8min & 20200301
\\ \hline
\cellcolor{Yellow} 20 & \cellcolor{Yellow} F & NY & \cellcolor{Yellow} 10 & 3.2min & \cellcolor{Yellow} 20200301
\\ \hline
40 & M & NY & 20 & 6.3min & 20200302
\\ \hline
\end{tabular}
\caption{A time series of relational data}
\label{fig:motivation}
\end{figure}}

\begin{figure*}[ht]
\hspace{-6cm}\begin{minipage}[b]{0.69\linewidth}
\footnotesize\center
\begin{tabular}{|c|c|c||c|c||c|}
\hline
\cellcolor{brown}{\sf Age $\dima\supind{1}$} & \cellcolor{brown}{\sf Gender $\dima\supind{2}$} & \cellcolor{brown}{\sf Location $\dima\supind{3}$} & \cellcolor{lime}{\sf Impression $\meaa\supind{1}$} & \cellcolor{lime}{\sf ViewTime $\meaa\supind{2}$} & \cellcolor{cyan}{\sf TimeStamp $t$}
\\ \hline \hline
\cellcolor{yellow} 30 & \cellcolor{yellow} F & WA & \cellcolor{yellow} 5 & 1.6min & \cellcolor{yellow} 20200301
\\ \hline
60 & M & WA & 1 & 1.8min & 20200301
\\ \hline
\cellcolor{yellow} 20 & \cellcolor{yellow} F & NY & \cellcolor{yellow} 10 & 3.2min & \cellcolor{yellow} 20200301
\\ \hline
40 & M & NY & 20 & 6.3min & 20200302
\\ \hline
\end{tabular}
\vspace{-0.4cm}
\caption{A time series of relational data (yellow cells are relevant to the task)}
\label{fig:motivation}
\end{minipage}

\vspace{-2cm}
\begin{minipage}[t]{0.69\linewidth}
\centering
\includegraphics[width=0.9\linewidth]{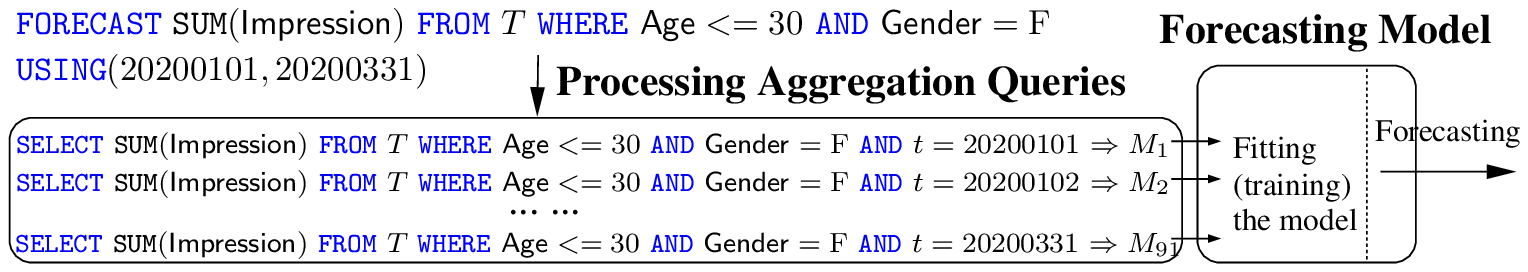}
\vspace{-0.4cm}
\caption{Processing a real-time forecasting task}
\label{fig:execution:example}
\vspace{-0.4cm}
\end{minipage}
\begin{minipage}[t]{0.3\linewidth}
\centering
\includegraphics[width=1\linewidth]{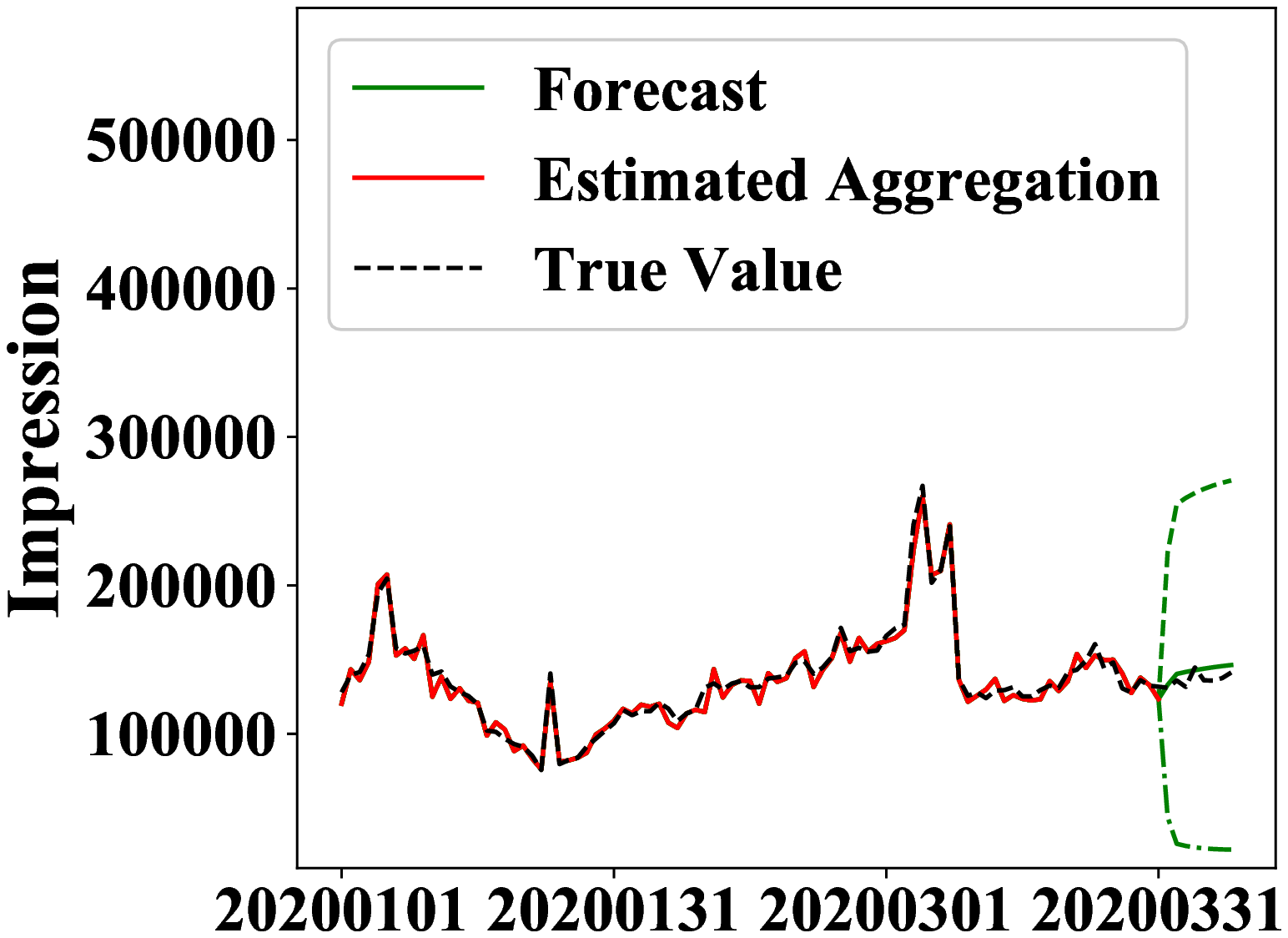}
\vspace{-0.8cm}
\caption{Forecasting example}
\label{fig:execution:result}
\end{minipage}
\vspace{-0.4cm}
\end{figure*}

\section{Introduction}
Large volumes of high-dimensional data are generated on eCommerce platforms every day, from data sources about, \eg, sales and browsing activities of online customers. 
Forecasting is among the most important BI analytical tasks to utilize such data, as sound prediction of future trends helps make informed business decisions.



\stitle{Motivating scenario and challenges.}
An online advertising platform enables advertisers to target certain user visits and submit their bids for displaying advertisements on these visits during a time window (\ie, {\em targeting ads campaigns}). 
To make the right decision about which customers to target and the bid prices, it is imperative for an advertiser to access reliable forecasts of important measures about the targeted customers and their activities, \eg, {\sf Impression}, the number of times an advertisement can be showed to such customers, and {\sf ViewTime}, the time a customer spends each visit.

Time-series data about customers can be very high-dimensional, with tens or hundreds attributes about customers' profiles and activities, including their demographics, devices (\eg, mobile/PC), machine-learned tags (\eg, their preferences and intents of visits), and so on. An advertiser may decide to target a group of customers for any combination of the attributes and values, based on the merchandise in the advertisement and fine-grained forecasts after different ways of slicing and dicing in the attribute space. For example, she may decide to target 20-30 year old females interested in sports and located in some cities for skirt sets in the advertisement.

Consider a time series of relation data in \cfig\ref{fig:motivation}. A forecasting task, \eg, the one in \cfig\ref{fig:execution:example}, asks to predict the total number of impressions by female customers under age 30.

Unlike in traditional forecasting applications such as airline planning, real-time response is critical in our scenario. First, it is an exploratory process to find the right attribute combination for targeting. An advertiser may try a number of combinations and read the forecasting results within a short period of time in order to support the decision. High latency can easily make advertisers impatient during the exploration. Second, real-time bidding for targeting ads campaigns is very competitive and dynamic market. Slow response may result in loss of displaying opportunities and higher prices. Thus, it is important to make our platform interactive.

It is challenging to provide real-time response to such forecasting tasks. The volume of input time-series data to the analytical pipeline is huge, with hundreds of millions of rows per day, and commonly months of historical data is used to forecast a future point.
Moreover, as a result of the high-dimensionality of our data, there could be trillions of possible attribute/value combinations; a forecasting task is given online with any combination, and it would be too expensive to precompute and store all possible combinations and the corresponding time series during an offline phase.

\stitle{The task and solution phases.}
Our forecasting system \sysname is designed to process {\em real-time forecasting tasks}, in which we specify: i) a constraint specifying the portion of data to be focused on (\eg, targeting customers ``${\sf Age}$ $<=$ $30$ $\code{AND}$ ${\sf Gender}$ $=$ ${\rm F}$'' in the example in \cfig\ref{fig:execution:example}); ii) the aggregated measure to be predicted (\eg, ${\tt SUM}({\sf Impression})$); iii) historical data points to be used to train a forecasting model (\eg, from $20200101$ to $20200331$). 
 
We give an overview of the solution.
Consider the example in \cfig\ref{fig:execution:example}. A forecasting task is specified in a SQL-like language, with the goal to predict the total number of impressions by female customers under age 30. To prepare training data for a forecasting model, we need to know the total number of impressions on each day ($M_t$), which can be written as an aggregation query. As the time-series data is partitioned on time, we can process these 91 aggregation queries with one scan of the data. After that, we have 91 data points to fit (train) the forecasting model. 
Processing the 91 aggregation queries (or, equivalently, a query with ${\tt \code{GROUP} ~ \code{BY}}$) is the bottleneck. A natural idea is to utilize sampling (\eg, \cite{jacm:DuffieldLT07}) or sample-based approximate query processing (\eg, \cite{sigmod:DingHCC016,sigmod:ParkMSW18}) to estimate their answers.
\cfig\ref{fig:execution:result} shows an example of prediction for the next 7 days: the red line shows estimated aggregations; the estimations are used to train the model, which then produces forecasts (green line) with confidence intervals (green dashed lines).
%


\stitle{Contributions and organization.}
\sysname is implemented in Alibaba's advertising system. 
%
%
The first technical question is how the errors in sample-based estimations affect the fitting of forecasting models (\csec\ref{sec:rtforecast}). Error bounds of estimating aggregations using samples are well studied \eg, for priority sampling \cite{stoc:Szegedy06} which has been shown to be optimal for aggregating $\tt SUM$. However, it is unclear what the implications of such error bounds are in prediction results. We provide both analytical and experimental evidence showing that aggregation errors add up to the forecasting model's noise (from historical data points), and they together decide how confident we could be with the prediction.
We give formal analysis for a concrete forecasting model (\ie, ${\rm ARMA}(1,1)$, defined later), and experimental results for more complicated models (\eg, LSTM).

The second question is about the space budget needed for samples (\csec\ref{sec:sampler}). 
Uniform sampling provides unbiased estimations for aggregations, but the estimation error is proportional to the range of a measure ($\max - \min$) \cite{sigmod:Hellerstein97}.
Weighted sampling schemes offer optimal estimations, with much better error bounds that are independent on the range \cite{stoc:Szegedy06}; however, as the sampling distribution is decided by the measure values \cite{jacm:DuffieldLT07,pods:AlonDLT05}, we have to draw one weighted sample per measure independently. When there are a number of measures (as in our scenario), the total space consumption could be prohibitive. We propose a new sampling scheme called \samplename ({\em \underline{G}eneralized \underline{S}moothed \underline{W}eighted}) sampling, with sampling distributions that can be arbitrarily specified. We analyze estimation error bounds by quantify the correlation between the sampling distribution and measure values. We introduce how to use this scheme to generate a compact sample which takes care of multiple measures, using a sampling distribution that averages distributions of different measures, and analyze its estimation error bounds.

We describe how \sysname is implemented in \csec\ref{sec:deploy}. We report experimental results on real datasets in \csec\ref{sec:exp}, and discuss related work in \csec\ref{sec:related}. 
\ifdefined\fullversion
	All missing proofs are deferred to \capp\ref{app:details}.
\else
	All missing proofs are in the full version \cite{url:tr}.
\fi


\section{Solution Overview}
\label{sec:overview}

%
\stitle{Time-series data model.} The input to our forecasting pipeline is a {\em time series of relation} $T$, \ie, a sequence of {\em observed rows} at successive time. We assume that time is a discrete variable here. A row in the table $T$ is specified by a pair $(i,t)$: an item $i$ and a time stamp $t$. An item is associated with multiple {\em dimensions}, each denoted by $\dima$, which are used to filter data (\eg, $\sf Age$ and $\sf Location$), and multiple {\em measures}, each denoted by $\meaa$, which we want to analyze and forecast (\eg, $\sf Impression$ and $\sf ViewTime$). We use $\dimv_{i,t}$ and $\meav_{i,t}$ to denote the values of a dimension and a measure, respectively, on each row; or, simply, $\dimv_{i}$ and $\meav_{i}$, if the time stamp $t$ is clear from the context or not important.
%
%
The schema of $T$ is $(\dima\supind{1}, \dima\supind{2}, \ldots, \dima\supind{d_a}; \meaa\supind{1}, \meaa\supind{2}, \ldots, \meaa\supind{d_m}; t)$.

\stitle{Real-time forecasting task.} A {\em forecasting tasks} is specified as:
\begin{align}
& \code{FORECAST} ~ {\tt SUM}(\meaa) ~ \code{FROM} ~ T ~ \code{WHERE} ~ \pred ~ \code{USING} ~ (t_s, t_e) \label{equ:task}
\\
& \code{OPTION} ~ ({\tt MODEL} = {\tt 'model\_x'}, {\tt FORE\_PERIOD} = {\tt t\_future}) \nonumber
\end{align}

Here, $\meaa$ is the measure we want to forecast from data source $T$ on a given set of rows satisfying the constraint $\pred$. \sysname allows $\pred$ to be any logical expression on the dimension values of $\dima\supind{1}, \ldots, \dima\supind{d_a}$. We want to use historical data from time stamp $t_s$ to $t_e$ to fit the forecasting model. We can also specify the forecasting model we want to use in the $\code{OPTION}$ clause with the parameter $\tt MODEL$, and the number of future time stamps we want to predict with the parameter $\tt FORE\_PERIOD$ (\eg, 7 days).
In most of our application scenarios, we care about $\tt SUM$ aggregation, \eg, total number of impressions from a certain group users; \sysname can also support $\tt COUNT$ and $\tt AVG$.

%
\sysname facilitates the following class of forecasting models.
Let $M_t$ be the value of metric $M$ ($={\tt SUM}(\meaa)$ in our case) at time $t$.
A model is specified by a time-dependent function $f_t$ with order $K$:
\begin{equation}\label{equ:forecastmodel}
M_t = f_t(M_{t-1}, M_{t-2}, \ldots, M_{t-K}).
\end{equation}
The model $f_t$ is {\em fitted} on historical data $M_1, M_2, \ldots, M_{t_0}$ with {\em training data tuples}, each in the form of $(M_{t}; M_{t-1}, \ldots, M_{t-K})$ with $M_{t-1}, \ldots, M_{t-K}$ as the input to $f_t$ and $M_t$ as the output, for $t = t_0, t_0-1, \ldots, K+1$. The fitted model can then be used to {\em forecast} future values $M_{t_0+1}, M_{t_0+2}, \ldots$ as $\hat M_{t_0+h|t_0}$ for $h=1, 2, \ldots$, iteratively (\ie, $\hat M_{t_0+1|t_0}$ can be used forecast $M_{t_0+2}$).

For the forecasting models supported by \sysname, we now give brief introduction to both classic ones, \eg, ARMA \cite{book:Hamilton94}, and those based on recurrent neural networks and LSTM \cite{neco:HochreiterS97}.

\stitle{Forecasting using ARMA.}
An ARMA ({\em autoregressive moving average}) \cite{book:Hamilton94} model
%
%
%
%
uses stochastic processes to model how $M_t$ is generated and evolves over time. 
It assumes that $M_t$ is a noisy linear combination of the previous $p$ values; each $u_t$ is an independent identically distributed zero-mean random noise at time $t$, and historical noise at previous $q$ time stamps impacts $M_t$ too:
%
%
%
%
%
%
%
%
\begin{equation} \label{equ:arma}
{\rm ARMA}(p,q): ~ M_t = \sum_{i=1}^p \alpha_i M_{t-i} + u_t + \sum_{i=1}^q \beta_i u_{t-i}.
\end{equation}
%

For example, an ARMA model is: $M_t = 0.8 M_{t-1} + 0.2 M_{t-2} + u_t + 0.1 u_{t-1}$. It falls into the form of \eqref{equ:forecastmodel}, and is parameterized by $\alpha_1 \ldots \alpha_p$ and $\beta_1 \ldots \beta_q$. The model can be fitted on $M_1, \ldots, M_{t_0}$, using, \eg, least squares regression, to find the values of $\alpha_i$ and $\beta_j$ which minimize the error term, and used to forecast future values.
%
%
%
We can estimate {\em forecast intervals}, \ie, {confidence intervals for forecasts} $\hat M_{t_0+h|t_0}$: with a {\em confidence level} (probability) $\gamma$, the true future value $M_{t_0+h}$ is within $[\hat M_{t_0+h|t_0} - l_\gamma, \hat M_{t_0+h|t_0} + r_\gamma]$. 

%
When there are deterministic trends over time, the {\em differential method} can be used: the first order difference is defined  $\triangledown M_t = M_t - M_{t-1}$, and the second order $\triangledown^2 M_t = \triangledown M_t - \triangledown M_{t-1}$, and in general, the $d$-th order $\triangledown^d M_t = \triangledown^{d-1} M_t - \triangledown^{d-1} M_{t-1}$.
If $\{\triangledown^d M_t\}_t$ is an ${\rm ARMA}(p,q)$ model, $\{M_t\}_t$ is an ${\rm ARIMA}(p,d,q)$ model. 

\stitle{Forecasting using LSTM.}
LSTM ({\em long short-term memory}) \cite{neco:HochreiterS97} is a network architecture that extends the memory of recurrent neural networks using a {\em cell}. It is natural to use LSTM to learn and memorize trends and patterns of time series for the purpose of forecasting. In a typical LSTM-based forecasting model, \eg, \cite{ijcai:SchmidhuberWG05}, an LSTM unit with dimensionality $d$ in the output space takes $(M_{t-1},$ $\ldots,$ $M_{t-K})$ as the input; the LSTM unit is then connected to a $d \times 1$ {\em fully-connected layer} which outputs the forecast of $M_t$.
This model also falls into the general form \eqref{equ:forecastmodel}. Here, the cell state is evolving over time and, at each time stamp $t$, is encoded in $f_t$, which can be learned from training data tuples $(M_{t};$ $M_{t-1},$ $\ldots,$ $M_{t-K})$ in order. 

\subsection{Overview of Our Approach}
Our system \sysname works in two online phases to process a forecasting task: {\em preparing training data points by issuing aggregation queries}, and {\em fitting the forecasting model using the training data}.

\squishlist
\item {\em Aggregation query.} In a forecasting task, specified in \eqref{equ:task}, to predict ${\tt SUM}(\meaa)$, we have $t_e - t_s + 1$ historical data points:
\begin{align}
M_{t_s} & = \code{SELECT} ~ {\tt SUM}(\meaa) ~ \code{FROM} ~ T ~ \code{WHERE} ~ \pred ~ \code{AND} ~ t = t_s \nonumber
\\
\ldots & \ldots \nonumber
\\
M_{t_e} & = \code{SELECT} ~ {\tt SUM}(\meaa) ~ \code{FROM} ~ T ~ \code{WHERE} ~ \pred ~ \code{AND} ~ t = t_e \label{equ:task:agg}
\end{align}
Each data point is given by an {\em aggregation query} with constraint $\pred$, which can be any logical expression on the dimension values of $\dima\supind{1},$ $\ldots,$ $\dima\supind{d_a}$. We would compute them in the online phase.
\item {\em Forecasting.} In the next online phase, we use $M_{t_s}, \ldots, M_{t_e}$ as training data to fit the forecasting model. And we use the model to predict future aggregations, $M_{t_e+1}, \ldots, M_{t_e + {\tt t\_future}}$.
\squishend

\stitle{Performance bottleneck.}
%
%
%
Suppose we have $N$ rows in $T$ for each time stamp, and use a history of $t_0 = t_e - t_s + 1$ time stamps to train a forecasting model with size (number of weights) $s$. The total cost of processing a forecasting task is $\bigohinline{t_0 \cdot N} + {\rm Train}(t_0, s)$, where $\bigohinline{t_0 \cdot N}$ is the cost of processing $t_0$ aggregation queries by scanning the table $T$, and ${\rm Train}(t_0, s)$ is the time needed to train a forecasting model with size $s$ using $t_0$ training data tuples. ${\rm Train}(t_0, s)$ is in the form of, \eg, $(t_0 \cdot s \cdot {\rm iter})$ where ${\rm iter}$ is the number of iterations for the model training to converge. 
We typically have $t_0$ in hundreds ($t_0 = 365$ if we use one year's history for training with one data point per day), $N$ (number rows in $T$ per day) in tens or hundreds of millions in our application, and $s$ in tens. Therefore, as $t_0, s \ll N$, processing of aggregation queries is the performance bottleneck (even in comprison to training a complex model).

\stitle{Real-time forecasting on approximate aggregations.}
In order to process a forecasting task in an interactive way in \sysname, we propose to estimate $M_{t_s},$ $\ldots,$ $M_{t_e}$ as $\hat M_{t_s},$ $\ldots,$ $\hat M_{t_e}$ from offline samples drawn from $T$, and use these estimates to form training data tuples and to fit the model \eqref{equ:forecastmodel}. There are several questions to be answered. i) If estimates $\hat M_i$, instead of $M_i$, are used to fit the forecasting model, how much the prediction would deviate. ii) How to draw these samples efficiently (preferably in a distributed manner). iii) How much space we need to store these samples for multiple measures.

\section{Real-Time Forecasting}
\label{sec:rtforecast}
%

We first analyze how sampling and approximate aggregations impact model fitting and, resultingly, forecasts. We give an analytical result for a special case of ARMA model, and will conduct experimental study for more complex models, \eg, LSTM-based ones in \csec\ref{sec:exp}. 
It is not surprising that there is a tradeoff between sampling rate (or, quality of estimated aggregations) and forecast accuracy.




\eat{
Consider an {\em autoregressive–moving-average} (ARMA) model
\begin{equation} 
{\rm ARMA}(p,q): ~ M_t = \sum_{i=1}^p \alpha_i M_{t-i} + u_t + \sum_{i=1}^q \beta_i u_{t-i}
\end{equation}
which is a combination of an {\em auto-regression} model ${\rm AR}(p):$
$M^{\rm AR}_t$ $=$ $\sum_{i=1}^p \alpha_i M_{t-i},$
and a {\em moving average} model ${\rm MA}(q):$
$M^{\rm MA}_t$ $=$ $u_t + \sum_{i=1}^q \beta_i u_{t-i},$
where each $u_t$ at time stamp $t$ is an independent zero-mean random noise. It is a stochastic process that models the evolving of the metric $M$ over time. After fixing $p$ and $q$, the model is fitted by, \eg, least squares regression, to find the values of $\alpha_1 \ldots \alpha_p$ and $\beta_1 \ldots \beta_q$ which minimize the error term.
}

\stitle{Required properties of sampling and estimates.}
%
In \sysname, we have no access to the accurate value of $M_t$; but instead, we have $\hat M_t$ estimated from offline samples. We require the estimates to have two essential properties for the fitting of forecast models:

%
%
\squishlist
\item ({\em Unbiasedness}) $\hat M_t = M_t + \eps_t$ with $\epinline{\eps_t} = 0$;
%
\item ({\em Independence}) $\eps_t$'s for different time stamps $t$ are independent.
\squishend
{\em \samplename sampling} introduced in \csec\ref{sec:sampler} offers unbiased estimates, with bounded variance of $\eps_t$. \samplename samplers are run independently on the data for each $t$---that is how we get independence.
%

%
\stitle{Impact on ${\rm ARMA}(p,q)$.}
When an ${\rm ARMA}(p,q)$ model has to be trained only on noisy metric values $\{\hat M_t\}$, we rewrite \eqref{equ:arma} as
%
\begin{align}
\hat M_t & = \sum_{i=1}^p \alpha_i \hat M_{t-i} + (u_t + \eps_t) + \label{equ:arma:noise}
\\
&
+ \begin{cases}
\sum_{i=1}^p (\beta_i u_{t-i} - \alpha_i \eps_{t-i}) + \sum_{i = p}^q \beta_i u_{t-i} & (1 < p \leq q)
\\
\sum_{i=1}^q (\beta_i u_{t-i} - \alpha_i \eps_{t-i}) - \sum_{i = q}^p \alpha_i \eps_{t-i} & (1 < q \leq p)
\end{cases} \nonumber
\end{align}
\eat{
\begin{align}
& \hbox{i) $1 < p \leq q$:} ~~~~ \hat M_t = \sum_{i=1}^p \alpha_i \hat M_{t-i} + \label{equ:arma:1}
\\
& ~~~~~~ + (u_t + \eps_t) + \sum_{i=1}^p (\beta_i u_{t-i} - \alpha_i \eps_{t-i}) + \sum_{i = p}^q \beta_i u_{t-i}. \label{equ:arma:1:noise}
\end{align}
\begin{align}
& \hbox{ii) $1 < q \leq p$:} ~~~~ \hat M_t = \sum_{i=1}^p \alpha_i \hat M_{t-i} + \label{equ:arma:2}
\\
& ~~~~~~ + (u_t + \eps_t) + \sum_{i=1}^q (\beta_i u_{t-i} - \alpha_i \eps_{t-i}) - \sum_{i = q}^p \alpha_i \eps_{t-i}. \label{equ:arma:2:noise}
\end{align}
}
In both cases, it is a combination of an autoregressive model \eqref{equ:arma:noise} of order $p$, and a moving average model of order $\max\{p,q\}$.
%
%
The model differs from the ARMA model in \eqref{equ:arma} only on the additional zero-mean error terms $\eps_t$, which are independent on other terms in the model and have known variance (for fixed sampling and the estimation methods). Thus, the model can be fitted on $\{\hat M_t\}$ using, \eg, maximum likelihood estimator, as normal ARMA models. $\eps_t$ increases the model's uncertainty, and, together with the model noise terms $u_t$, decides the confidence of the model prediction.

For a fixed confidence level, the narrower the forecast intervals are, the more confident we are about the prediction.
Their widths are proportional to the standard deviations of forecasts of $\hat M_t$, which in turn depends on the variance of noise terms $u_t$'s and $\eps_t$'s.

In comparison to the original ARMA model in \eqref{equ:arma}, the additional noise term $\eps_{t}$ will indeed incur wider forecast intervals.
However, with a proper sampling-estimation scheme, 
if $\eps_t$'s variance is negligible in comparison to $u_t$'s, $\eps_{t}$ will have little impact on the forecast error/interval, which will be demonstrated in our experiments later. Here, we give a formal analysis for the ${\rm ARMA(1,1)}$ model:
\begin{proposition}\label{prop:arma11}
{Suppose} we have a time series $\{M_t\}$ satisfies ${\rm ARMA}(1,1)$. Let $\hat M_t = M_t + \eps_t$ be an estimation of $M_t$ satisfying unbiasedness and independence. Then, $\vrinline{\hat M_t} = a \cdot \sigma_u^2 + \sigma_\eps^2$, where $\sigma_u^2 = \vr{u_t}$, $\sigma_\eps^2 = \vr{\eps_t}$, and {$a = (1+2\alpha_1\beta_1+\beta_1^2)/(1 - \alpha_1^2)$} is a constant decided by parameters in ${\rm ARMA}(1,1)$.
\end{proposition}
%
%
%
We can use normal random variables to approximate $u_t$ and $\eps_t$, and estimate forecast intervals for a given confidence level.

\eat{
In particular, when we use the optimal Bernoulli sampling to estimate $M_t$, we have $\vrinline{\eps_t} = \Delta M_t$ and
\begin{corollary}
If we use optimal Bernoulli sampling to estimate $M_t = \sum_{i=1}^n m_i$ as $\hat M_t$, with sampling parameter $\Delta$, under the condition that $m_i = \bigthetainline{M_t/n}$, the noise term $\eps_t = \hat M_t - M_t$ converges in distribution to a normal random variable:
\[
\frac{1}{\sqrt{\Delta M_t}}\eps_t \xrightarrow{d} N(0, 1).
\]
\end{corollary}
The proof is directly from our analysis of the optimal Bernoulli sampling in \csec\ref{sec:sampler:opt} and Lyapunov Central Limit Theorem. 
}

\begin{figure}[t]
\centering
\includegraphics[width=0.9\linewidth]{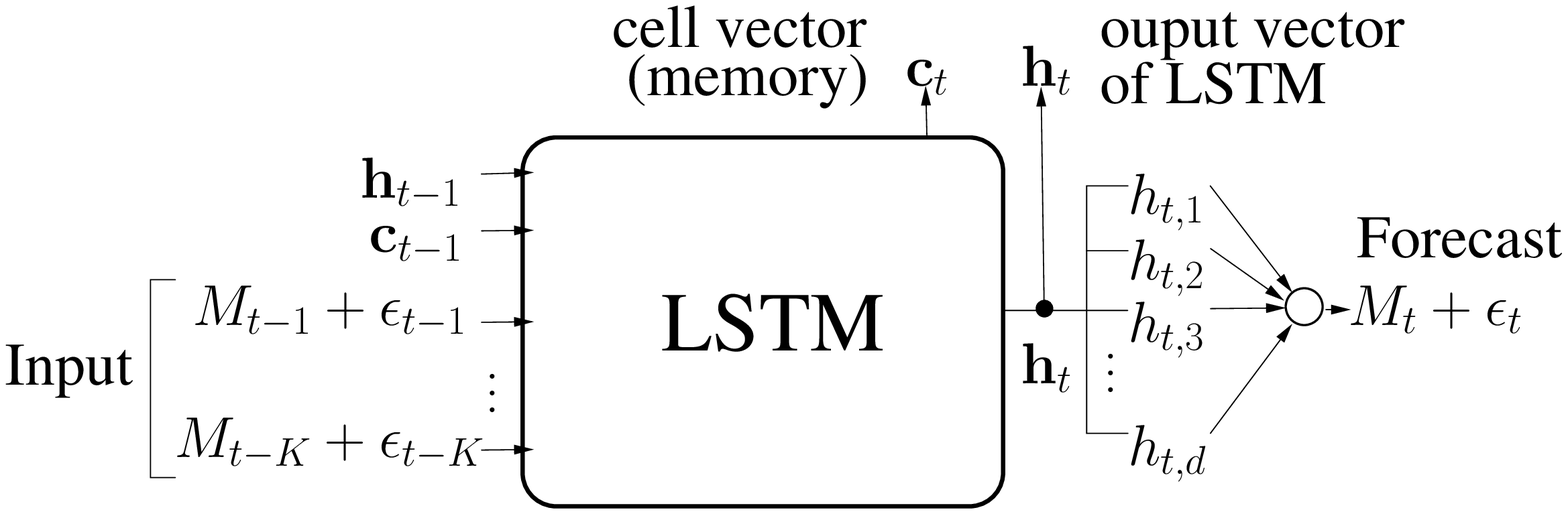}
\vspace{-0.4cm}
\caption{An LSTM-based forecasting model with noisy inputs}
\label{fig:lstm:noisy}
\vspace{-0.6cm}
\end{figure}

\stitle{Impact on LSTM-based model.} LSTM can be naturally applied for forecasting tasks thanks to its ability to memorize trends and patterns of time series. \cfig\ref{fig:lstm:noisy} depicts such a forecasting model and illustrates where noise in the estimates impacts the model fitting.

At time stamp $t$, we want to forecast $M_t$ with the previous $K$ metric values $M_{t-1}, \ldots, M_{t-K}$ and the ``memory'' $({\bf c}_{t-1}, {\bf h}_{t-1})$. However, we have only estimates $\hat M_{t-i} = M_{t-i} + \eps_{t-i}$ available; these estimates are fed into the LSTM unit as inputs. LSTM then generates an {\em output vector} ${\bf h}_t$ and update its memory {\em cell} from ${\bf c}_{t-1}$ to ${\bf c}_t$. ${\bf h}_t$ can be interpreted as the current status of LSTM and it is used as the input to a fully-connected layer for forecasting $M_t$. Again, we have only $\hat M_t = M_t + \eps_t$ available as an approximation, which we learn to fit with LSTM and the fully-connected layer. ${\bf c}_t$ and ${\bf h}_t$ are used to deliver memory to the next time stamp.

Noise terms $\eps_t$'s may affect how the weight matrices in LSTM and the fully-connected layer are learned and the values of ${\bf c}_t$ and ${\bf h}_t$ derived (via linear transformations and activation functions).
We conjecture that the LSTM-based forecasting model performs well as long as the estimates $\hat M_t$'s are accurate enough (or $\eps_t$'s are small enough). It is difficult to derive any formal analytical result here, but we will evaluate it experimentally in \csec\ref{sec:exp}. 
%
%

\section{Generalized Weighted Sampler}
\label{sec:sampler}


We now focus on how to draw samples for estimating results of aggregation queries. An aggregation query in \eqref{equ:task:agg} is essentially to compute the sum of a subset of measure values in a relation $T$ at time $t$; the subset is decided online by the constraint $\pred$ specified in the forecasting task. W.l.o.g., suppose the subset of rows satisfying $\pred$ is $[n] = \{1,\ldots,n\}$, we want to estimate the metric $M = \sum_{i=1}^n m_i$, for a measure $\meaa = [\meav_i]$. The (offline) sampling algorithm should be independent on $\pred$, but could depend on $\meaa$.

\stitle{Existing samplers.}
There are two categories of sampling schemes for estimating subset sums: {\em uniform sampling} and {\em weighted sampling}.
In uniform sampling, each row in the relation is drawn into the sample with equal probability; an unbiased estimation for $M$ is simply the rescaled sum of values in the sample. It has been used in online aggregation extensively, but the main deficiency is that the error bound is proportional to the difference between the maximum and minimum (or, the {\em range} of) measure values \cite{sigmod:Hellerstein97}.

In weighted sampling, each row $i$ is drawn with probability proportional to $m_i$, so that we can remove the dependency of the estimation's error bound on the range of the measure values. More concretely, \cite{sigmod:chaudhuri99} and \cite{sigmod:DingHCC016} give efficient implementations of such a sampling distribution: for some fixed constant $\tau$ (deciding the sampling size), if $m_i < \tau$, the probability of drawing a row $i$ is $m_i/\tau$, and if $m_i \geq \tau$, multiple (roughly $\tau/m_i$) copies of $i$ are included into the sample.
Threshold sampling \cite{tit:DuffieldLT05} and priority sampling \cite{pods:AlonDLT05,jacm:DuffieldLT07} differ from the above one in that, if $m_i \geq \tau$, exactly one copy of $i$ is included. Priority sampling has been shown to be optimal \cite{stoc:Szegedy06} in terms of the sampling efficiency with relative standard deviation $\sqrt{\vrinline{\rm estimation}}/M = \sqrt{1/({\rm sample\_size - 1}})$.

\stitle{Requirements for the sampler in \sysname.} 
Weighted sampling offers much better sampling efficiency than uniform sampling, especially for heavy tailed distributions, which is common in practice. However, in all the above weighted sampling schemes, the sampling distribution is decided by the measure's values. In \sysname, we have $d_m$ different measures to forecast; and thus, we have to draw $d_m$ such weighted samples independently; when $d_m$ is large (\eg, $10$), the storage cost of all the samples (\eg, even with sampling rate $1\%$) is prohibitive in memory. 
Therefore, the requirement here is: {\em whether we can generate a compact sample which takes care of multiple measures and still have accuracy guarantees}.

To this end, we propose {\em \samplename} ({\em \underline{G}eneralized \underline{S}moothed \underline{W}eighted}) sampling. We introduce its sampling distribution and analyze its sampling efficiency in \csec\ref{sec:sampler:gb}. We utilizes the ``generality'' of its sampling distribution to generate a compact sample which takes care of multiple measures, and analyzes under which conditions it gives provable accuracy guarantees in \csec\ref{sec:sampler:compact}.


\subsection{Generalized Smoothed Sampling}
\label{sec:sampler:gb}
%
%
%
The \samplename sampling scheme is parameterized by a positive constant $\Delta$ and positive sampling weights ${\bf w} = [w_i]$: each row $i$ in the given relation $T$ is drawn into the sample $S_\Delta$ with probability $\frac{w_i}{\Delta + w_i}$, {\em independently}. For fixed sampling weights, $\Delta$ decides the sample size; and the choice of $\bf w$ decides the estimation accuracy.


%
Inspired by the Horvitz–Thompson estimator \cite{jasa:HorvitzT52}, we define the {\em calibrated measure} to be $\hat m_i = m_i(\Delta + w_i)/w_i$ if row $i$ is drawn into the sample $S_\Delta$, and $\hat m_i = 0$ otherwise. We would associate $\hat m_i$ with each sample row $i$ and store it in $S_\Delta$. Formally, we have
\begin{equation}\label{equ:gbin:variable:probability}
\begin{cases}
\prinline{i \in S_\Delta \wedge \hat m_i = \frac{m_i(\Delta + w_i)}{w_i}} = \frac{w_i}{\Delta + w_i}
\\
\prinline{i \notin S_\Delta \wedge \hat m_i = 0} = 1 - \frac{w_i}{\Delta + w_i} = \frac{\Delta}{\Delta + w_i}
\end{cases}.
\end{equation}
%
%
We can estimate $M$ as $\hat M = \sum_{i \in S_\Delta} \hat m_i$, which is unbiased since
\[
\ep{{\hat M}} = \sum_{i=1}^n \ep{\hat m_i} = \sum_{i=1}^n \frac{m_i(\Delta + w_i)}{w_i} \cdot \frac{w_i}{\Delta + w_i} = M.
\]


\stitle{Simple and efficient implementations.} 
A \samplename sampler can be easily implemented in a distributed manner: each row $i$ generates uniformly random number $p_i$ from $[0,1]$, independently; according to \eqref{equ:gbin:variable:probability}, if $p_i \leq \frac{w_i}{\Delta + w_i}$, the row $i$ is put into the sample $S_\Delta$.

A \samplename sample $S_\Delta$ can be maintained in an incremental way. Suppose we have drawn $S_\Delta$ from rows $[n]$ for some fixed $\Delta$. If more rows $n+1, \ldots, n+k$ are coming, we want to increase $\Delta$ to $\Delta'$ and obtain a \samplename sample $S_{\Delta'}$ from $[n+k]$ with size comparable to $|S_\Delta|$. Suppose rows in $S_\Delta$ are sorted by $(\frac{1}{p_i} - 1)w_i$ in an ascending order; we only need to delete those with $\Delta \leq (\frac{1}{p_i} - 1)w_i < \Delta'$
from $S_\Delta$, and insert those with $\Delta' \leq (\frac{1}{p_i} - 1)w_i$, for $i = n+1, \ldots, n+k$. In this way, we update $S_\Delta$ to $S_{\Delta'}$ {\em without} touching any row in $[n] - S_\Delta$.

\subsubsection{Accuracy Guarantee on Aggregations}
\label{sec:sampler:gb:acc}
We now analyze estimation errors of the class of \samplename-based estimators $\hat M$, instantiated by $(\Delta,{\bf w})$.
%
%
%
We consider {\em relative standard deviation} ($\rm RSTD$) and {\em relative error} ($\rm RE$). As $\hat M$ is unbiased,
\begin{align*}
{\rm RSTD}({\hat M}) \! \triangleq \!\! \sqrt{\ep{\left(\frac{{\hat M}-M}{M}\right)^2}}
\geq \ep{\frac{|{\hat M} - M|}{M}} \!\! \triangleq \! {\rm RE}({\hat M}).
\end{align*}

The goal of our analysis is to establish a relationship between the sample size and the (expected) $\rm RSTD$ and $\rm RE$. Intuitively, when the sampling weight $w_i$ is ``consistent'' with the measure $m_i$, the estimation error is minimum (for fixed sample size). We introduce the following notation to quantify the ``consistency''.
\begin{definition}\label{def:consistency}
{\bf ($(\lowerb,\upperb)$-consistency)}
Sampling weights ${\bf w}$ $=$ $[w_i]$ are $(\lowerb,\upperb)$-consistent with measure $\meaa = [\meav_i]$, iff $\lowerb = \min_i (m_i/w_i)$ and $\upperb = \max_i (m_i/w_i)$. $\theta \triangleq \upperb/\lowerb$ is called the consistency scale.
\end{definition}

The above notation about ``consistency'' says, for any row $i$, $\lowerb \leq m_i/w_i \leq \upperb$. It allows $\meaa$ and ${\bf w}$ to differ in scale (\eg, both $\lowerb$ and $\upperb$ could be large) but measures their similarity in patterns and trends. For example, suppose $\meaa$ $=$ $[100,$ $100,$ $200,$ $400]$ and ${\bf w}$ $=$ $[10,$ $10,$ $20,$ $50]$. We have $\lowerb = 400/50 = 8$, $\upperb = 10$, and thus $\theta = 10/8 = 1.25$.
In general, we have $\theta \geq 1$, and the following theorem shows that, the relative error has an upper bound that is proportional to $\sqrt{\theta}$.
\begin{theorem}\label{thm:agg} {\bf (Sampling efficiency of \samplename)}
Suppose the sampling weights $\bf w$ used in \samplename sampling are $(\lowerb,\upperb)$-consistent with measure values $\bf m$, the estimate $\hat M$ is unbiased and has expected relative standard deviation and error bounded by: ($\theta \triangleq \upperb/\lowerb$)
\[
{\rm RE}({\hat M}) \leq {\rm RSTD}({\hat M}) \leq \sqrt{\frac{\upperb/\lowerb}{\ep{|{\mathcal S}_{\Delta}|}}} = \sqrt{\frac{\theta}{\ep{|{\mathcal S}_{\Delta}|}}}.
\]
\end{theorem}
%

An open question raised by Alon \etal when introducing priority sampling \cite{pods:AlonDLT05} is: whether we can provide any error bound if subset sum on a measure ($\meaa$) is estimated using a priority sample drawn based on a different measure ($\bf w$). \cthm\ref{thm:agg} answers this question in \samplename sampling by specifying under what condition ($(\lowerb,\upperb)$-consistency)) there is an error bound. 
%

\subsubsection{A Special Case: Optimal GSW Sampler}
\label{sec:sampler:opt}
We can choose $\bf w$ to minimize the variance of estimation, while the expected sample size is bounded.
\ifdefined\fullversion
	Please refer to \capp\ref{app:optweight} for a precise formulation and the optimal solution.
\else
	Please refer to \cite{url:tr} for a precise formulation and the optimal solution.
\fi
%
%
From \cthm\ref{thm:agg}, a nearly-optimal solution is ${\bf w} = \meaa$, as in this case, $\bf w$ is $(1,1)$-consistent with $\bf m$ ($\theta = 1$). We call it {\em optimal \samplename sampler}. Directly from \cthm\ref{thm:agg}, we have
%
%
\begin{corollary}\label{cor:agg:opt}
{\bf (Optimal \samplename sampler)} If we use ${\bf w} = {\bf m}$ ($w_i = m_i$ for each row $i$) as the sampling weights, we have
\[
{\rm RE}({\hat M}) \leq {\rm RSTD}({\hat M}) \leq  \sqrt{\frac{1}{\ep{|{\mathcal S}_{\Delta}|}}}.
\]
\end{corollary}

The optimal \samplename sampler has sampling efficiency that is comparable to the best known sampler for estimating subset sums, \eg, priority sampling \cite{pods:AlonDLT05} with ${\rm RSTD} = \sqrt{1/({\rm sample\_size - 1})}$. We will compare their empirical performance in \csec\ref{sec:exp}.

\subsection{Compact Sample for Multiple Measures}
\label{sec:sampler:compact}
The size of \samplename sample can be controlled by the parameter $\Delta$. When there is only one measure $\meaa$, we draw an optimal \samplename sample (setting ${\bf w} = \meaa$). A more common scenario is that we have multiple measures (\eg, in \cfig\ref{fig:motivation}) in one relation. We can draw one optimal \samplename sample per each measure, which, however, increases the space consumption significantly.
%
%
The question is whether we can use one sample to take care multiple measures. 

Suppose there are $k$ measures in a relation to be aggregated and predicted, $\meaa\supind{1},$ $\ldots,$ $\meaa\supind{k}$. For each measure $j$, we want to estimate $M\supind{j} = \sum_{i=1}^n \meav\supind{j}_i$ for rows in a set $[n]$ (satisfying the constraint $\pred$ in a forecasting task).
A \samplename sample $S_\Delta$ is drawn using weights ${\bf w} = [w_i]$.
%
%
The {\em calibrated measure} on each sample row $i$ for each measure $j$ is $\hat m\supind{j}_i = m\supind{j}_i(\Delta + w_i)/w_i$. From \cthm\ref{thm:agg}, $\hat M\supind{j} = \sum_{i \in S_\Delta} \hat m\supind{j}_i$ is an unbiased estimation of $M\supind{j}$.

Intuitively, if the chosen sampling weight vector $\bf w$ centers around $\meaa\supind{1},$ $\ldots,$ $\meaa\supind{k}$ and is not too far away from any of the $k$, from \cthm\ref{thm:agg}, the error can be better bounded. We can find such centers by taking the average of measures. For example, for $\meaa\supind{1} = [100, 100, 200, 400]$ and $\meaa\supind{2} = [1, 1, 2, 1]$, the {\em geometric mean} is ${\bf w}^\times = [\sqrt{100 \cdot 1} = 10, 10, 20, 20]$, and the {\em arithmetic mean} is ${\bf w}^+ = [(100 + 1)/2 = 50.5, 50.5, 101, 200.5]$.
We now analyze how the error can be bounded for these two choices.
%

\stitle{Geometric compressed \samplename sampling.} We can use the geometric mean of the $k$ measures as the sampling weights:
\begin{equation}
w^\times_i = \left(\prod_{j=1}^k \meav\supind{j}_{i}\right)^{1/k}.
\end{equation}

Among the $k$ measures to be grouped, define the {\em trend deviation} between any two measures $\meaa\supind{p}$, $\meaa\supind{q}$ (for $p,q \in [k]$):
\begin{equation}
\overline\rho_{p,q} \triangleq \max_{i\in[n]} \frac{\meav\supind{p}_{i}}{\meav\supind{q}_{i}},
~~~
\underline\rho_{p,q} \triangleq \min_{i\in[n]} \frac{\meav\supind{p}_{i}}{\meav\supind{q}_{i}},
~~~ \hbox{and} ~ \rho_{p,q} \triangleq \frac{\overline\rho_{p,q}}{\underline\rho_{p,q}},
\end{equation}
which measures how similar the trends (instead of their absolute values) of the two measures are\footnote{Or, equivalently, $\meaa\supind{q}$ is $(\underline\rho_{p,q}, \overline\rho_{p,q})$-consistent with $\meaa\supind{p}$.}. The smaller $\rho_{p,q}$ is, the more similar $\meaa\supind{p}$ and $\meaa\supind{q}$ are. For example, if $\meaa\supind{p} = c \cdot \meaa\supind{q}$ for a constant $c$, $\rho_{p,q} = c/c = 1$. Define $\rho \triangleq \max_{p, q \in [k]} \rho_{p,q}$ to be the {\em maximum deviation} among the $k$ measures. From \cthm\ref{thm:agg},



%
\begin{corollary}\label{cor:agg:geo}
If we use ${\bf w}^\times = [w^\times_i]$ as the sampling weights for a relation with $k$ measures, with a \samplename sample $S_\Delta$, we can estimate $M\supind{p}$ as $\hat M\supind{p}$ for each measure $p$ with error 
\[
{\rm RE}({\hat M}\supind{p}) \leq {\rm RSTD}({\hat M}\supind{p}) \leq \sqrt{\frac{\prod_{j: ~ j \neq p}\rho_{p,j}^{1/k}}{\ep{|{\mathcal S}_{\Delta}|}}} \leq \sqrt{\frac{\rho^{\frac{k-1}{k}}}{\ep{|{\mathcal S}_{\Delta}|}}}.
\]
\end{corollary}
%


\stitle{Arithmetic compressed \samplename sampling.}
Another choice is to use the arithmetic mean as the sampling weights:
\begin{equation}
w^+_i = \frac{1}{k} \sum_{j=1}^k \meav\supind{j}_{i}.
\end{equation}

Define the {\em range deviation} $\delta$ among $k$ measures: for each row $i$, consider the ratio between the maximum measure and the minimum one; $\delta$ is the maximum ratio among all rows:
\begin{equation}
\delta \triangleq \max_{i\in[n]}\left(\frac{\max_{j\in[k]} \meav\supind{j}_{i}}{\min_{j\in[k]} \meav\supind{j}_{i}}\right).
\end{equation}

From the definitions, for any measure $p$, we have $1/\delta$ $\leq$ $\meav\supind{p}_{i}/w^+_i$ $\leq$ $\delta$. Thus, directly from \cthm\ref{thm:agg}, we have the following bound.
\begin{corollary}\label{cor:agg:ari}
If we use ${\bf w}^+ = [w^+_i]$ as the sampling weights for a relation with $k$ measures, with a \samplename sample $S_\Delta$, we can estimate $M\supind{p}$ as $\hat M\supind{p}$ for each measure $p$ with error 
\[
{\rm RE}({\hat M}\supind{p}) \leq {\rm RSTD}({\hat M}\supind{p}) \leq \sqrt{\frac{\delta^2}{\ep{|{\mathcal S}_{\Delta}|}}}.
\]
\end{corollary}
%

\stitle{How to group measures?}
In the above, we have shown that, for chosen sampling weights, how the error can be bounded with some parameters ($\rho$ and $\delta$) decided by the data about a group of measures.
When there are many (\eg, $k=100$) measures, $\rho$ and $\delta$ could be huge and thus the above error bounds are not informative. We want to partition measures into small groups of size 4-5 based on their correlation, so that within each group one \samplename sample gives good estimations for the measures. To this end, we establish a relationship between $(\lowerb,\upperb)$-consistency and $L_1$ distance as follows.
\begin{proposition}\label{prop:consistencyl2}
{\bf (consistency and $L_1$ distance)}
Suppose sampling weights ${\bf w}$ $=$ $[w_i]$ are $(\lowerb,\upperb)$-consistent with measure $\meaa = [\meav_i]$. We normalize $\bf w$ as ${\bf w}' = [w'_i = w_i/\sum_j w_j]_i$, and similarly, $\meaa$ as $\meaa'$. Let $\theta = \upperb/\lowerb$. We have $\|\meaa' - {\bf w}'\|_1 \leq (\theta - 1)$.
\end{proposition}

$(\lowerb,\upperb)$-consistency is a ``worst-case'' notion about all the rows. It is not a good metric for grouping measures, because first, it is expensive to compute $\theta$, and second, an aggregation query may not touch all the rows.
Instead, we use $L_1$ distance (after normalization) as in \cprop\ref{prop:consistencyl2} to quantify the {\em correlation} between two measures. Intuitively, it quantifies how two measures are similar in trends and patterns regardless of their absolute values. 

We consider a formulation based on the {{\sc KCenter} problem} \cite{book:Harpeled2011}. The goal is to partition the measures into $g$ groups so that the max $L_1$ distance (after normalization) between any measure to the center of the group it belongs to is minimized. 
Here, the $L_1$ distance between two measures can be estimated using a sample of rows. We can apply the standard greedy algorithm \cite{book:Harpeled2011} to find a 2-approximation.
Within each group, we use the geometric/arithmetic mean as the sampling weight vector. 
It is a heuristic strategy without any formal guarantee, but from \cprop\ref{prop:consistencyl2} and the triangle inequality in $L_1$, at least we know that we'd better not group two measures that are far way (\eg, $> 2(\theta-1)$) together, as there is no sampling weight vector that is consistent with both of them at the same time.
%
%
\ifdefined\fullversion
	We conduct a preliminary evaluation on the above correlation metric, $L_1$ distance (after normalization). Using the same datasets and workloads (with sensitivity ranging from $0.5\%$ to $10\%$) as Exp-I in \csec\ref{sec:exp}, the experimental results are plotted in \cfig\ref{fig:exp:correlation}. Here, the four measures, {\sf Impression}, {\sf Click}, {\sf Favorite}, and {\sf Cart}, are partitioned into two equal-size groups (there are a total of three ways of partitioning). For example, the first way in \cfig\ref{fig:exp:correlation} is group1 = \{{\sf Impression}, {\sf Click}\} and group2 = \{{\sf Favorite}, {\sf Cart}\}. For each group, we pick the sampling weight vector as the arithmetic mean of measure vectors in this group. \cfig\ref{fig:exp:correlation:l1} plots the $L_1$ distance between each measure vector and the sampling weight vector, and \cfig\ref{fig:exp:correlation:aerr} plots the aggregation error using this sampling weight vector in \samplename. It can be seen that aggregation errors and $L_1$ distances have similar trends. That is, if a sampling weight vector is close to a measure vector, it gives better estimation; thus, if two measure vectors are close in $L_1$, it is better to put them in one group (their mean can be used as the sampling weight vector).
\fi
\ifdefined\fullversion
	In \csec\ref{sec:exp}, we focus on experimental evaluation after the grouping is fixed. We will leave the development and analysis of better grouping strategies as future work.
\else
	A preliminary evaluation of this grouping strategy can be found in \cite{url:tr}.
	%
\fi

\ifdefined\fullversion
\begin{figure}[t]
\begin{minipage}[t]{0.46\linewidth}
\subfigure[Three ways to group the four measures: {\sf Imp} = {\sf Impression}, {\sf clk} = {\sf Click}, {\sf fav} = {\sf Favorite}, {\sf cart} = {\sf Cart}]{
\centering
\includegraphics[width=1\linewidth]{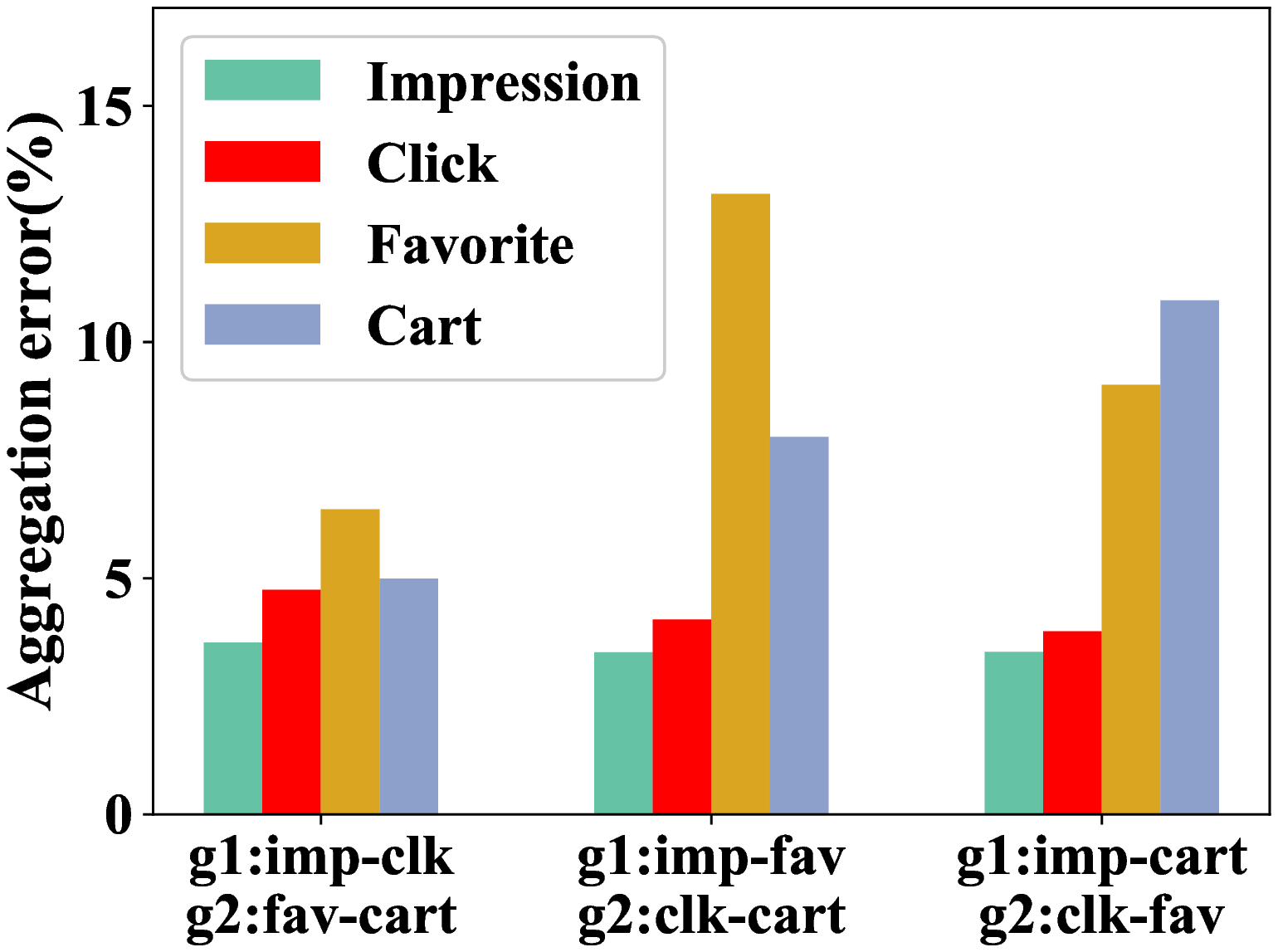}
\label{fig:exp:correlation:aerr}
}
\end{minipage}
\hspace{0.26cm}
\begin{minipage}[t]{0.46\linewidth}
\subfigure[$L_1$ distance between measure vectors and the sampling weight vector (the arithmetic mean of measure vectors in a group)]{
\centering
\includegraphics[width=1\linewidth]{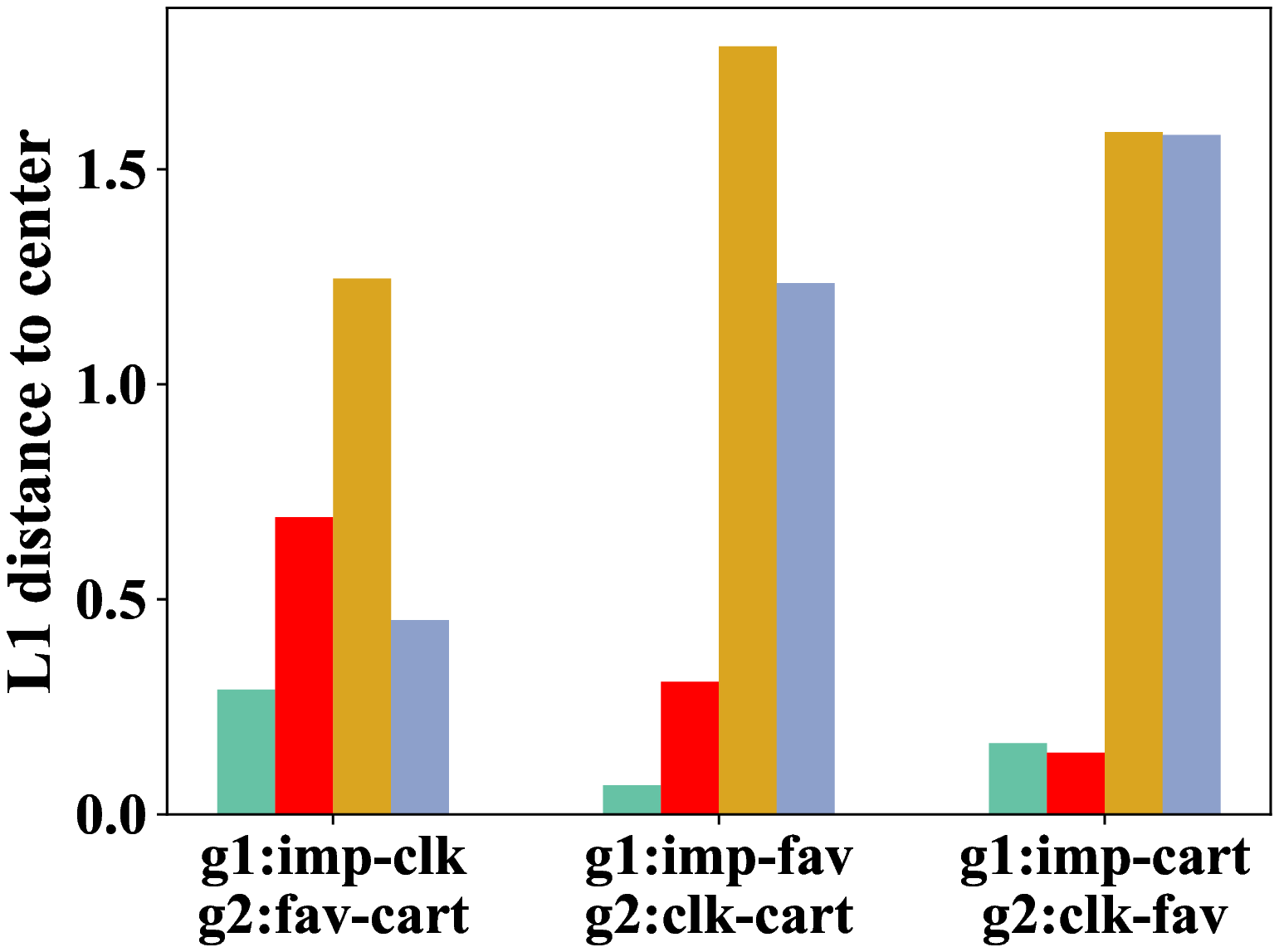}
\label{fig:exp:correlation:l1}
}
\end{minipage}
\vspace{-0.5cm}
\caption{Aggregation error and $L_1$ distance}
\vspace{-0.3cm}
\label{fig:exp:correlation}
\end{figure}
\fi

\ifdefined\fullversion
\begin{figure}[t]
\centering
\includegraphics[width=1\linewidth]{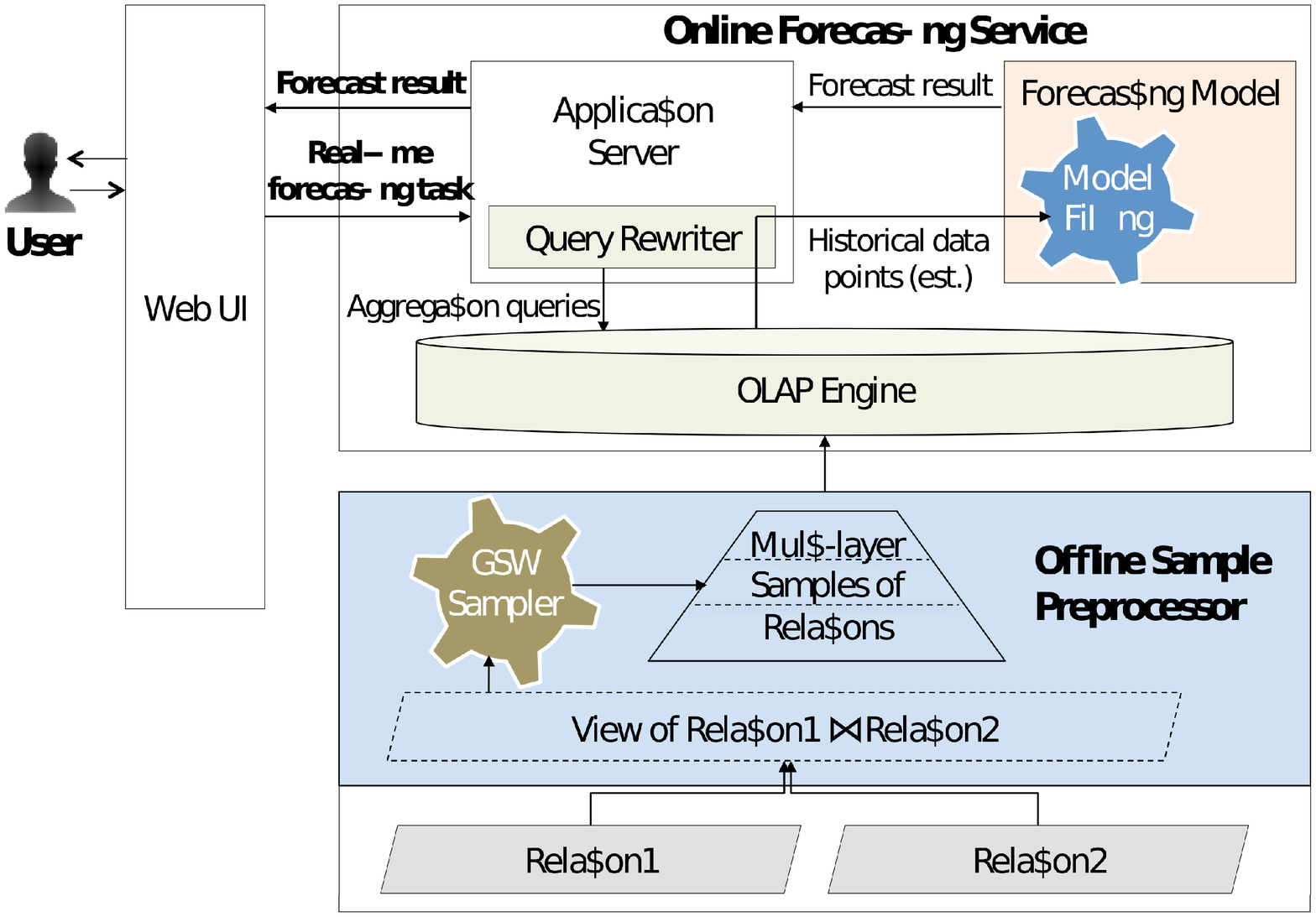}
\vspace{-1.0cm}
\caption{Deployment of the system \sysname}
\vspace{-0.4cm}
\label{fig:deploy}
\end{figure}
\fi

\section{Deployment}
\label{sec:deploy}
\ifdefined\fullversion
We have implemented and deployed \sysname in Alibaba's ads analytical platform.
\cfig\ref{fig:deploy} shows important components of \sysname.
\fi

\stitle{Offline Sample Preprocessor} of \sysname is built on Alibaba's distributed data storage and analytics service, MaxCompute \cite{url:alibabacloud}. Time series of relations, partitioned by time, are stored in MaxCompute's data warehouse. Relations can be joined, \eg, tables $\sf UserProfile$ and $\sf AdTraffic$ are joined on $\sf UserID$. GSW sampler is implemented as UDFs (user-defined functions), and draws samples from one relation or the view of joined relation. A set of samples of different sizes (with increasing values of $\Delta$) are drawn and stored as Multi-layer Samples of Relations for response time-prediction accuracy tradeoff.

\stitle{Online Forecasting Service} first pulls Multi-layer Samples of Relations into Alibaba's in-memory OLAP engine, Hologres \cite{url:alibabacloud}, to enable real-time response. 
There are 30 servers in the cluster to support this service, each with 96 CPU cores and 512G memory.
Sample data is partitioned by time in the OLAP engine.

Users submit forecasting tasks to an Application Server via a Web UI. For a task, aggregation queries in \eqref{equ:task:agg} are generated from Query Rewriter and processed on samples in the OLAP engine to obtain estimated answers. These estimations are used as training data to fit the Forecasting Model, which is built on a Python server. 

Two models are implemented. One is ARIMA. An open-source library \cite{url:autoarima} (built on X-13ARIMA-SEATS \cite{url:x13arima}) that trains ARIMA and automatically tunes for the best values of parameters $p,d,q$ is used. We also support an LSTM-based model (in \cfig\ref{fig:lstm:noisy}), which is implemented using Keras \cite{url:keras}. We use the LSTM and fully-connected layers in Keras with $K = 7$ and $d = 4$ as the default parameter setting.
%
%
Other forecasting models can be plugged in here, too. 
After fitting the model, we send forecasts back to users.

\section{Experimental Evaluation}
\label{sec:exp}
We evaluate our system \sysname under the implementation and hardware specified in \csec\ref{sec:deploy},
on a real-life dataset 
%
%
with 11 dimensions about users' profiles
and 4 numeric measures to be predicted, including $\sf Favorite$, $\sf Impression$, $\sf Click$, and $\sf Cart$. There are around 15 million rows per day, and 200 days of data.
%

We compare different samplers: {\bf Uniform} sampling, which is also used in \cite{sigmod:AgarwalCLSV10}, is the baseline; {\bf Priority}, the optimal weighted sampler \cite{jacm:DuffieldLT07}; our {\bf Optimal \samplename} and {\bf Arithmetic/Geometric compressed \samplename} introduced in \csec\ref{sec:sampler}. We also compare our sampling based methods with {\bf PIM} (Partwise Independence Model) \cite{sigmod:AgarwalCLSV10} based on a Bayesian model assuming partially independence.

In the following, forecasting tasks are randomly picked with different measures to be predicted and different combinations of dimensions in their constraints, with some (approximately) fixed {\em selectivity} (the fraction of rows satisfying the constraint). By default, we use 150 days' data to fit the model and predict the next 7 days; we report {\em relative aggregation errors} (average of the 150 days), {\em relative forecast error}, and {\em forecast intervals} (average of the next 7 days), taking the average of 400 independent runs of different tasks, together with one standard deviation, for each measure and for each value of selectivity (on independent samples).

\stitle{Exp-I: A summary of results.}
We first give a brief summary of experimental results. \ctab\ref{fig:exp:methods} reports the average forecast errors on 20 random tasks with selectivity from $0.5\%$ to $10\%$ using ARIMA. ``Full'' stands for the result when we use the full data to process aggregation queries for training. With a sampling rate $0.1\%$, our optimal \samplename and compressed \samplename in \sysname perform consistently better than Uniform and PIM in terms of forecast errors, and sometimes are very close to Full (the best we can do). These two also offer interactive response time (less than $100$ms). In the rest part, we will report more detailed results about response time and performance of different sampling-based methods.

\begin{table}[t]
\small\centering
\caption{{A summary of results (0.1\% sample, Opt-GSW = Optimal GSW, C-GSW = Arithmetic compressed GSW)}}
\label{fig:exp:methods}
\vspace{-0.4cm}
\begin{tabular}{|c|c|c|c|c|c|}
\hline
 & Full & PIM & Uniform & Opt-GSW & C-GSW
\\ \hline \hline
$\sf Favorite$ & 0.105  & 0.695 & 0.248 & 0.131 & 0.196
\\ \hline
$\sf Impression$ & 0.140 & 0.374 & 0.147 & 0.142 & 0.144
\\ \hline
$\sf Click$ & 0.157 & 0.681 & 0.161 & 0.151 & 0.153
\\ \hline
$\sf Cart$ & 0.704 & 1.931 & 0.718 & 0.704 & 0.709
\\ \hline
\end{tabular}
\vspace{-0.4cm}
\end{table}

\begin{wrapfigure}{r}{0.18\textwidth}
 	\vspace{-0.6cm}
 	\begin{center}
 		\hspace{-0.9cm}\includegraphics[width=0.23\textwidth]{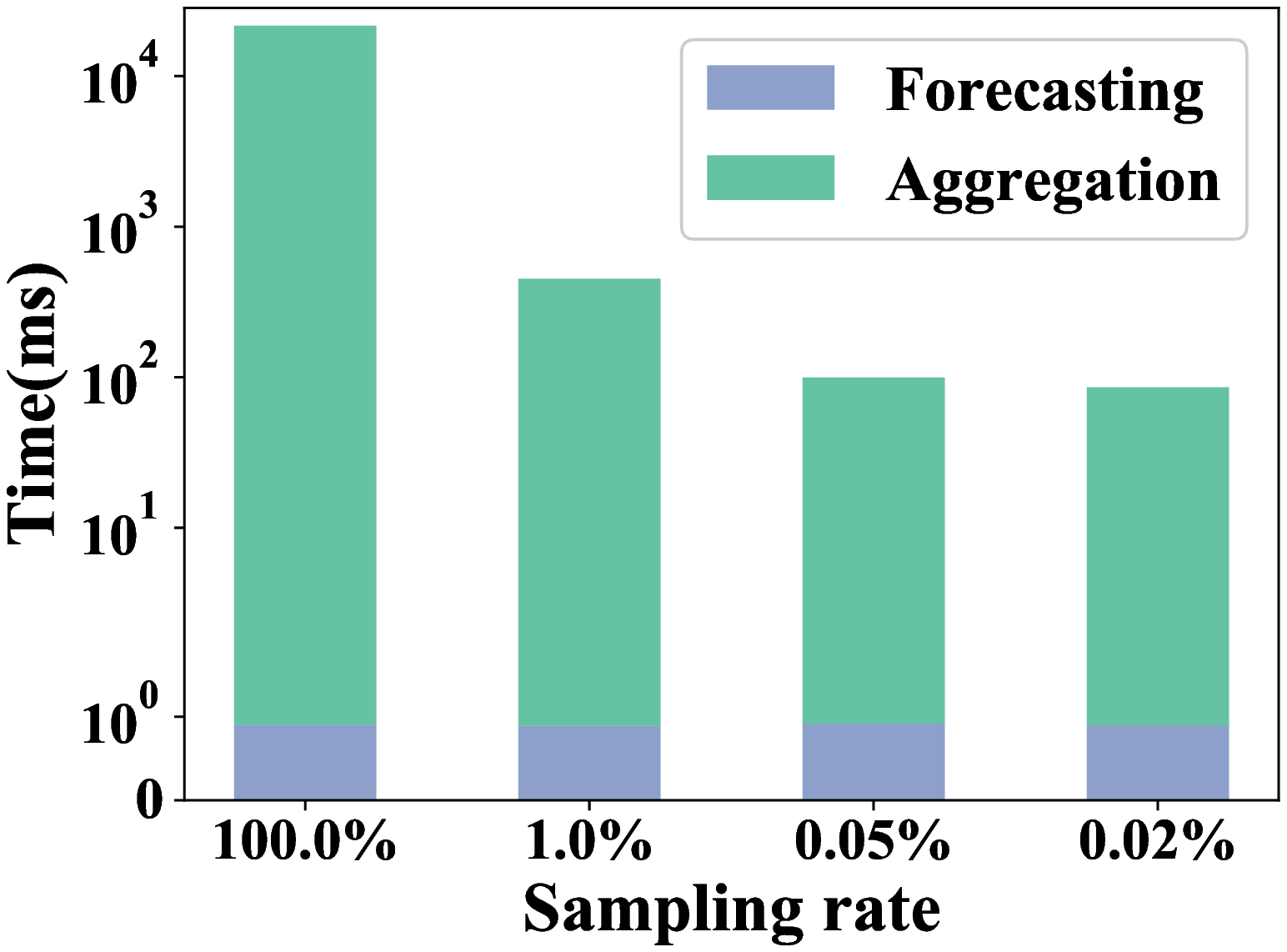}
 	\end{center}
 	\vspace{-0.4cm}
 	\caption{E2E response time (with ARIMA)}
 	\label{fig:exp:scalability:time:05}
 	\vspace{-0.4cm}
\end{wrapfigure}

\stitle{{Exp-II: Real-time response.}}
The end-to-end response time is reported in \cfig\ref{fig:exp:scalability:time:05}, partitioned into the portion for processing (estimating) aggregation queries, and the portion for forecasting (model fitting + prediction using ARIMA). It can be seen that the portion for aggregation queries is the bottleneck, but with sampling, the response time can be reduced from around 20sec on the full data to 30ms on a $0.02\%$ sample, which still gives reasonable prediction as will be shown later. If LSTM is used, the model fitting is much more expensive, but we still have an interactive response time around 1sec if a $1\%$ sample is used.

\ifdefined\fullversion
\stitle{Exp-III: Varying number of time stamps used in training.}
We consider different numbers of training data points used to fit the model. We focus on tasks with selectivity $5\%$ to forecast {\sf Impression}. The average forecast errors, with one standard deviation, for varying sampling rate using our optimal \samplename are plotted in \cfig\ref{fig:exp:scalability}. It shows that the number of time stamps used to fit model has an obvious impact on the forecast error, with 150 (days) giving the most accurate and stable prediction for both ARIMA and LSTM. It motivates us to speedup the processing of aggregation queries, as more time stamps mean more aggregation queries.
%
%
%
We also test different selectivities on other measures. The trends are similar.
\else
\stitle{Exp-III: Varying number of time stamps used in training.}
We consider different numbers of training data points used to fit the ARIMA and LSTM models. Please refer to the full version \cite{url:tr} for more details.
%
%
It shows that the number of time stamps used to fit model has an obvious impact on the forecast error, with 150 (days) giving the most accurate and stable prediction for both ARIMA and LSTM. It motivates us to speedup the processing of aggregation queries, as more time stamps mean more aggregation queries.
%
%
%
\fi

\ifdefined\fullversion
\begin{figure}[ht]
\vspace{-0.2cm}
\begin{minipage}[t]{0.479\linewidth}
\subfigure[ARIMA]{
\centering
\includegraphics[width=1\linewidth]{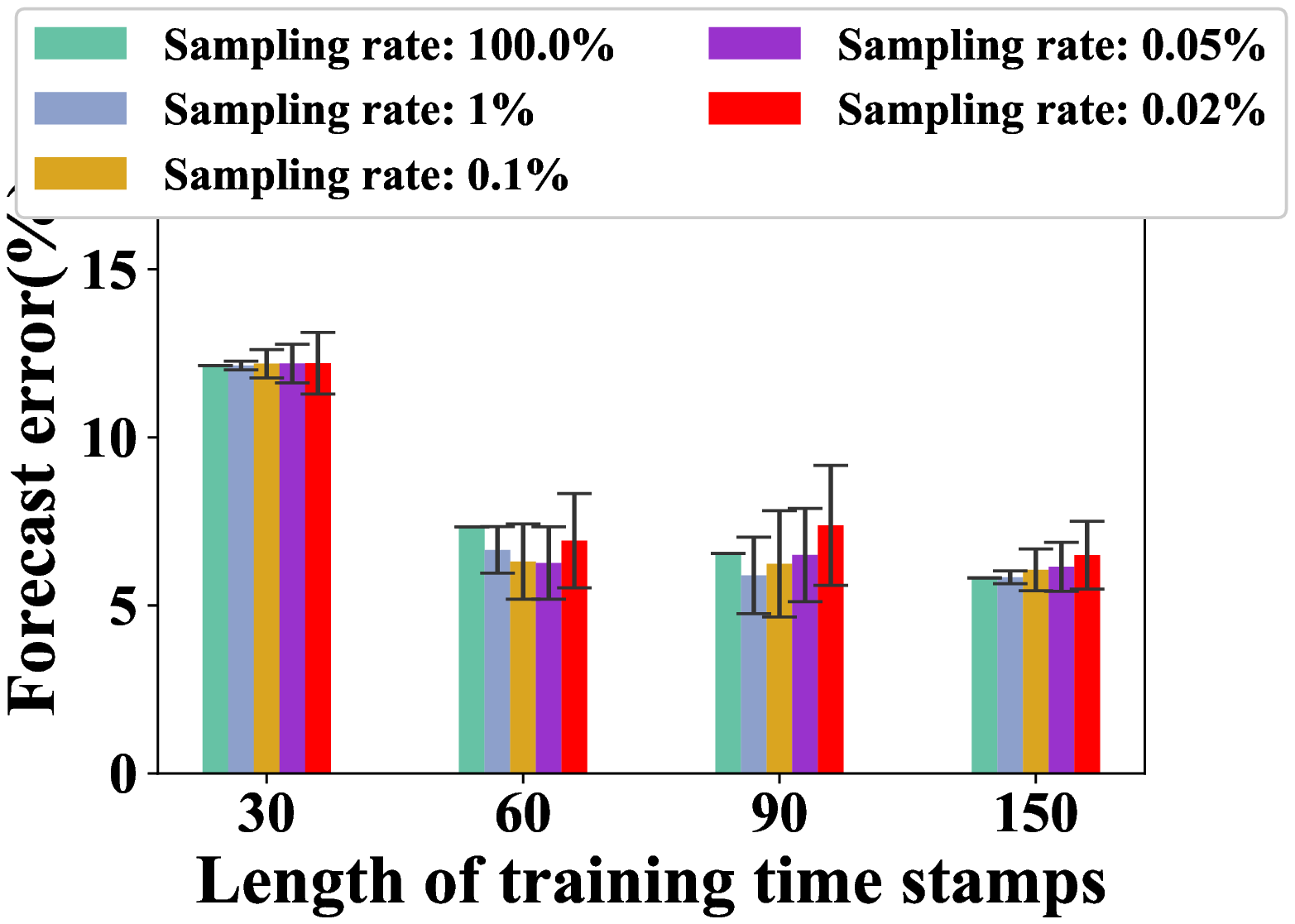}
\label{fig:exp:scalability:error:05}
}
\end{minipage}
~~~
\begin{minipage}[t]{0.479\linewidth}
\subfigure[LSTM]{
\centering
\includegraphics[width=1\linewidth]{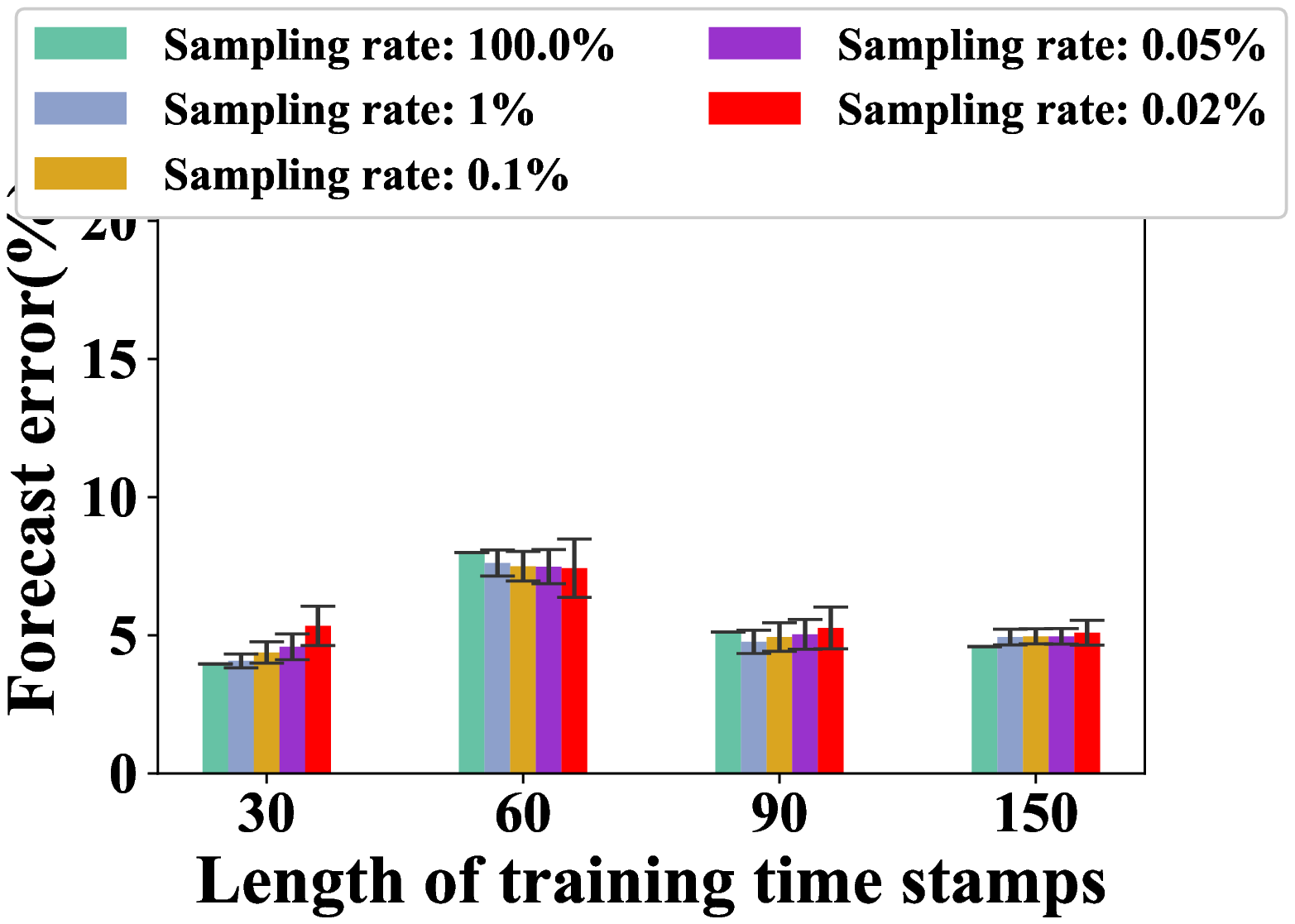}
\label{fig:exp:scalability:error:05:lstm}
}
\end{minipage}
\vspace{-0.5cm}
\caption{Number of time stamps used in training v.s. Forecast error (selectivity $5\%$, $\sf Impression$)}
\vspace{-0.3cm}
\label{fig:exp:scalability}
\end{figure}
\fi

\begin{figure}[ht]
\vspace{-0.2cm}
\subfigure[Selectivity 0.5\%]{
\begin{minipage}[t]{0.47\linewidth}
\centering
\includegraphics[width=1.1\linewidth]{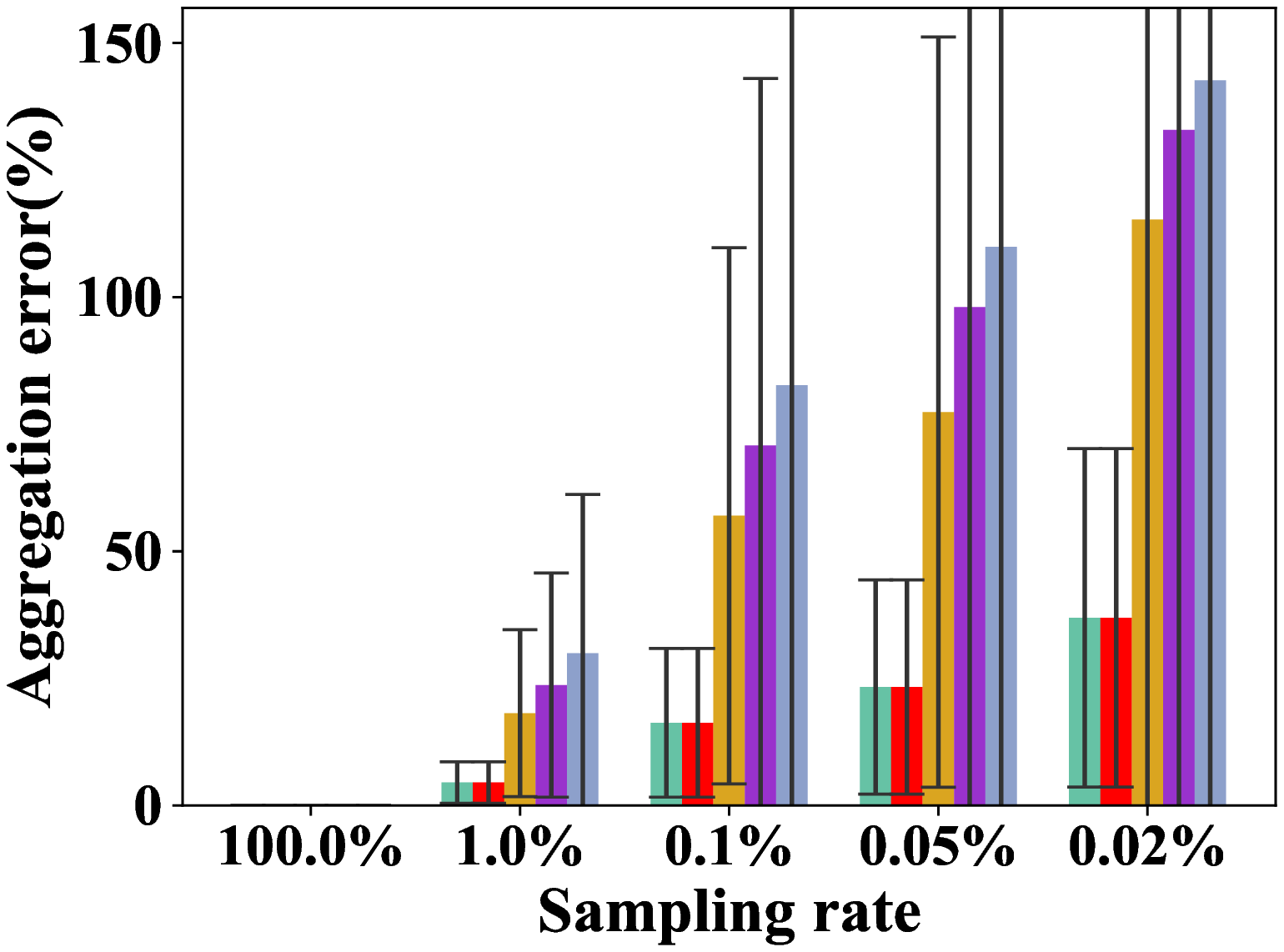}
\end{minipage}
}
\subfigure[Selectivity 5\%]{
\begin{minipage}[t]{0.47\linewidth}
\centering
\includegraphics[width=1.1\linewidth]{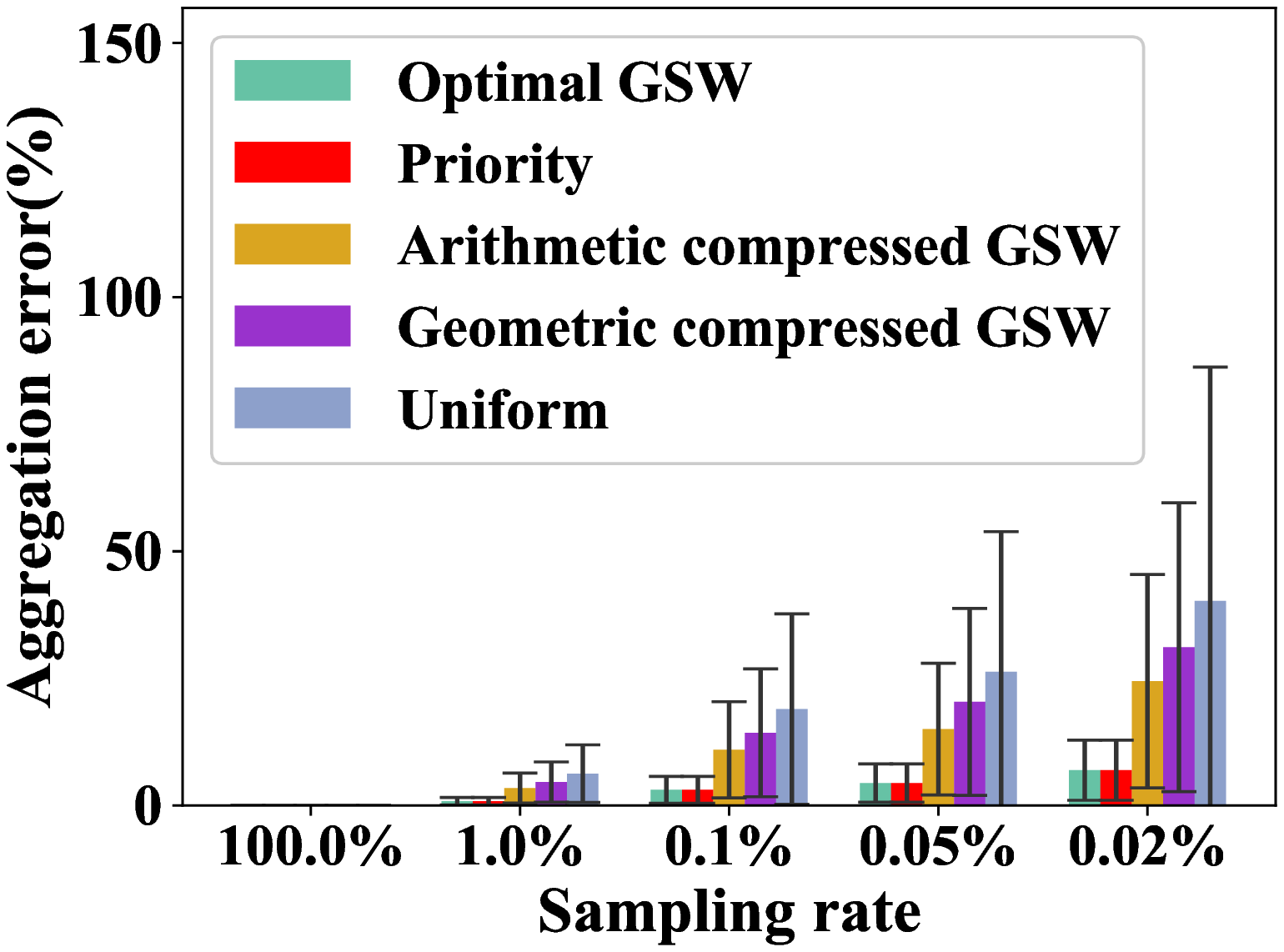}
\end{minipage}
}
\vspace{-0.3cm}
\caption{Aggregation error of different sampling methods for varying selectivity and sampling rate ({\sf Favorite})}
\vspace{-0.2cm}
\label{fig:exp:aerror:favorite}
\end{figure}

\begin{figure}[t]
\vspace{-0.2cm}
\subfigure[ARIMA]{\label{fig:exp:ferror:favorite:arima}
\begin{minipage}[t]{0.47\linewidth}
\centering
\includegraphics[width=1.1\linewidth]{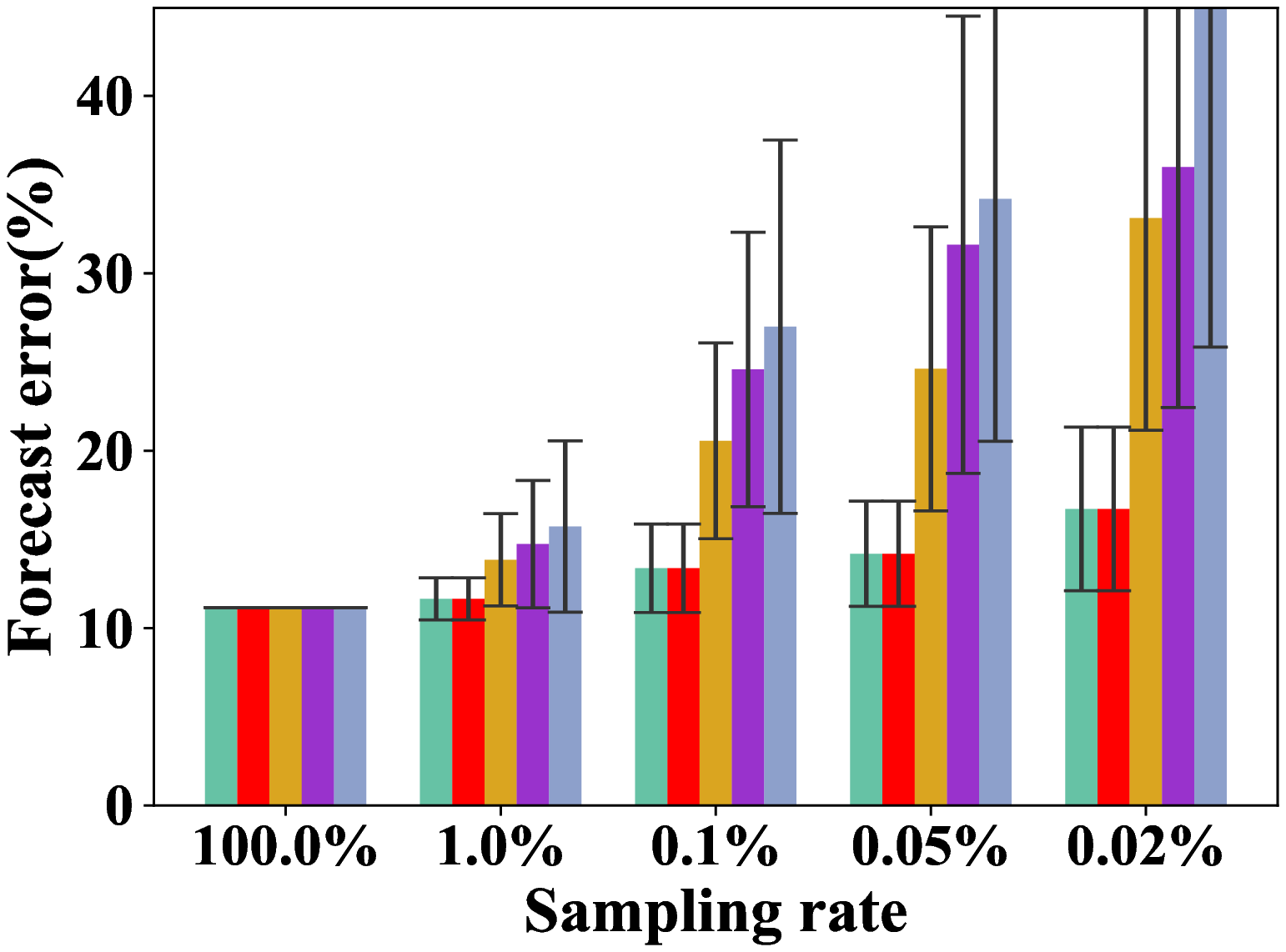}
\end{minipage}
}
\subfigure[LSTM]{\label{fig:exp:ferror:favorite:lstm}
\begin{minipage}[t]{0.47\linewidth}
\centering
\includegraphics[width=1.1\linewidth]{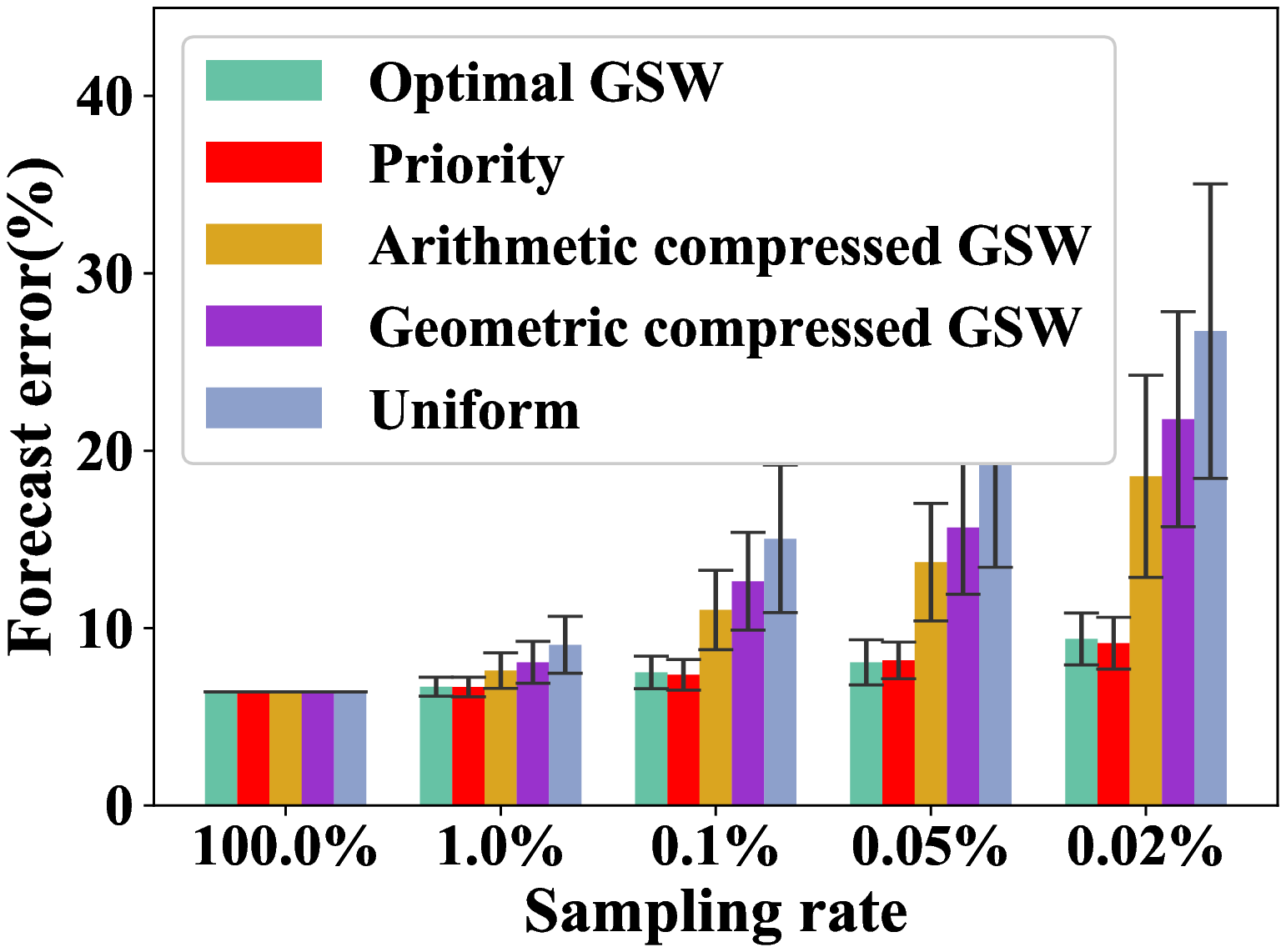}
\end{minipage}
}
\vspace{-0.3cm}
\caption{Forecast error of different sampling methods for varying sampling rate (selectivity 0.5\%, {\sf Favorite})}
\vspace{-0.4cm}
\label{fig:exp:ferror:favorite}
\end{figure}

\ifdefined\fullversion
\begin{figure}[t]
\vspace{-0.2cm}
\subfigure[ARIMA]{\label{fig:exp:ferror:favorite:5:arima}
\begin{minipage}[t]{0.47\linewidth}
\centering
\includegraphics[width=1.1\linewidth]{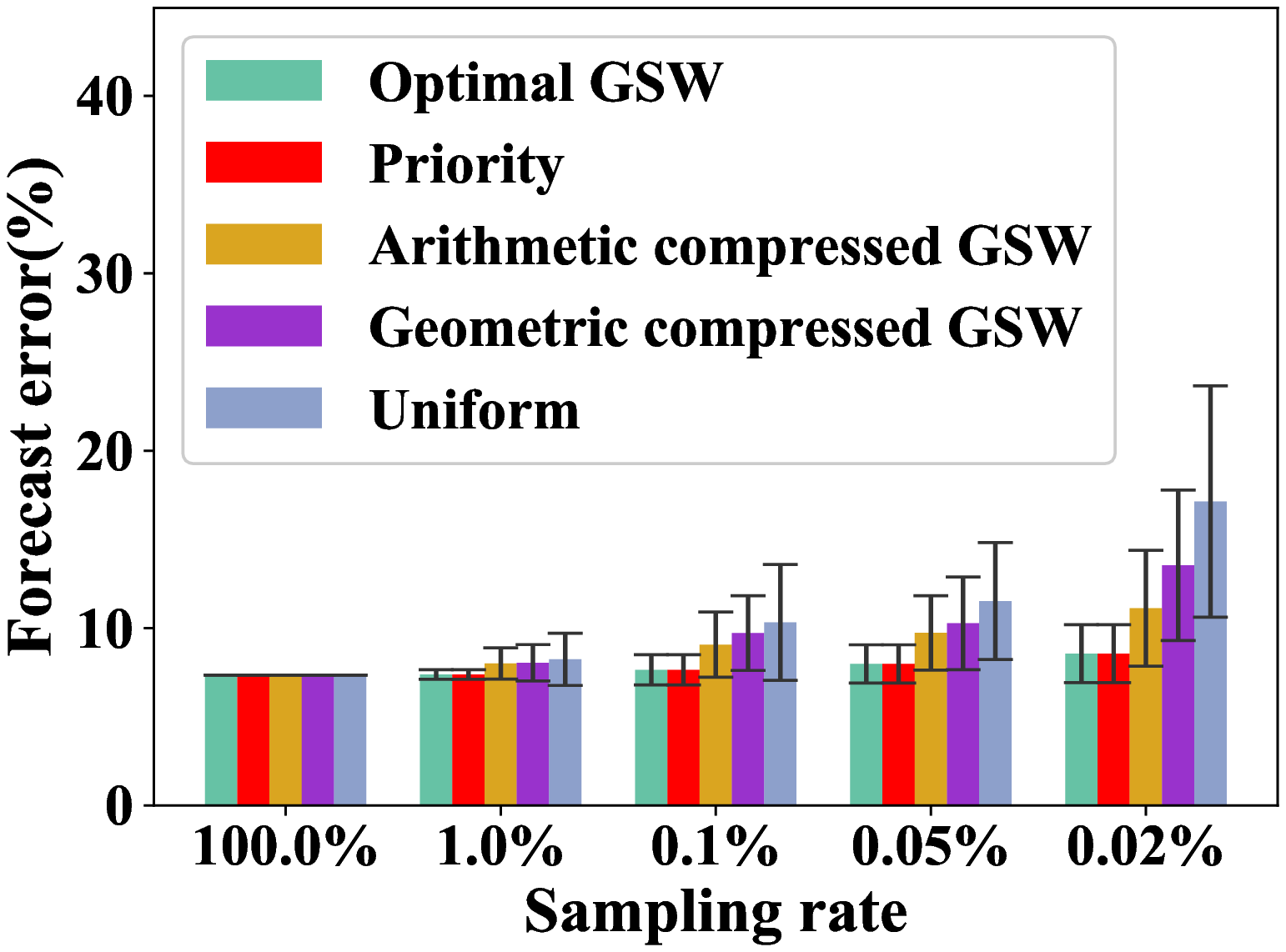}
\end{minipage}
}
%
\subfigure[LSTM]{\label{fig:exp:ferror:favorite:5:lstm}
\begin{minipage}[t]{0.47\linewidth}
\centering
\includegraphics[width=1.1\linewidth]{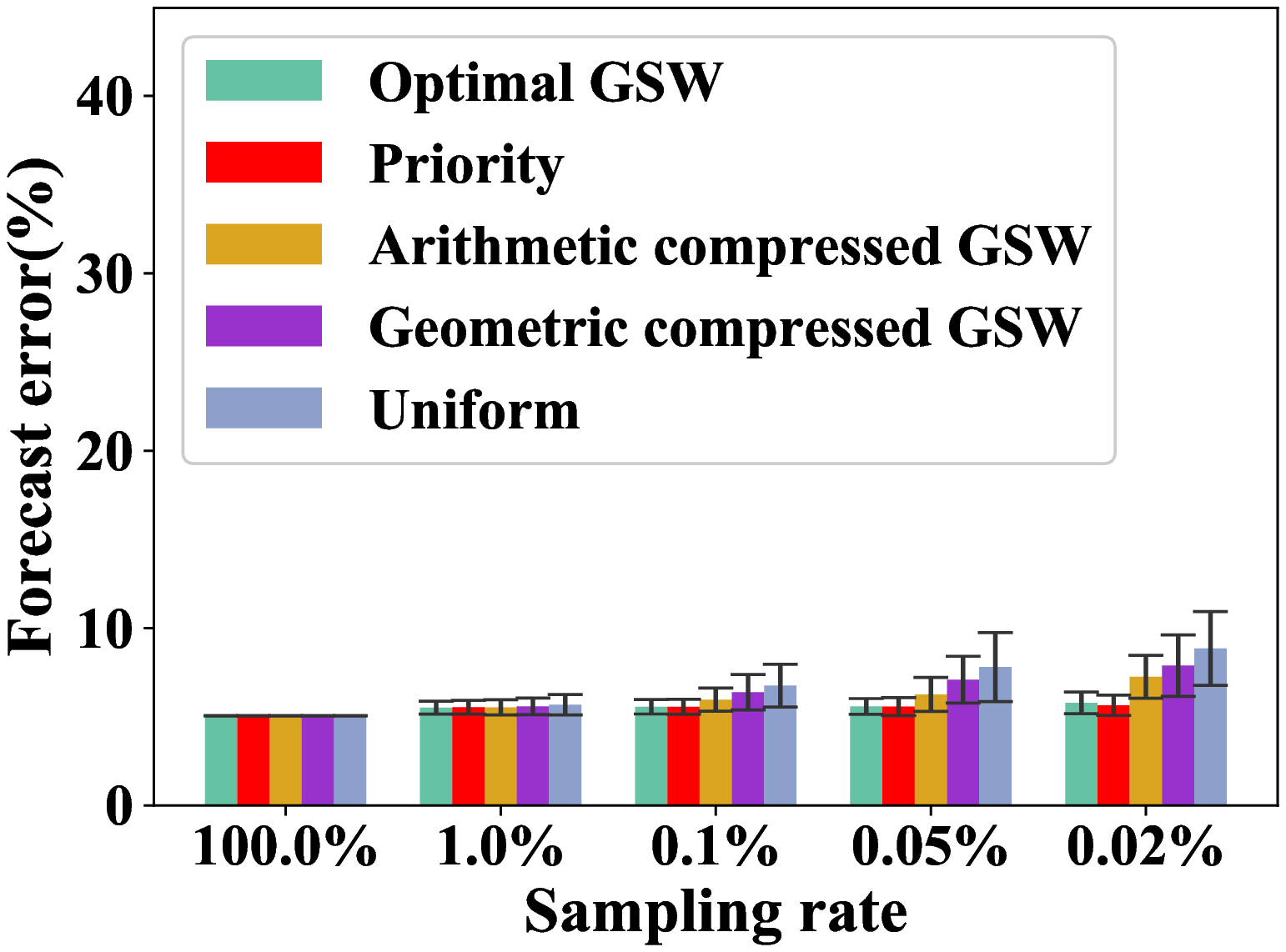}
\end{minipage}
}
\vspace{-0.3cm}
\caption{Forecast error of different sampling methods for varying sampling rate (selectivity 5\%, {\sf Favorite})}
\vspace{-0.4cm}
\label{fig:exp:ferror:favorite:5}
\end{figure}
\fi

\stitle{Exp-IV: Varying sampling rate and selectivity.} We compare different samplers in \sysname, for varying sampling rate and selectivity. Note that for Arithmetic/Geometric compressed \samplename, one sample suffices for the relation; Priority are Optimal \samplename are measure-dependent, and thus using either of them we need four samples (one per measure), with the total space consumption four times of Uniform and compressed \samplename for a fixed sampling rate.

\begin{figure}[t]
\vspace{-0.2cm}
\subfigure[ARIMA]{\label{fig:exp:ferror:impression:arima}
\begin{minipage}[t]{0.47\linewidth}
\centering
\includegraphics[width=1.1\linewidth]{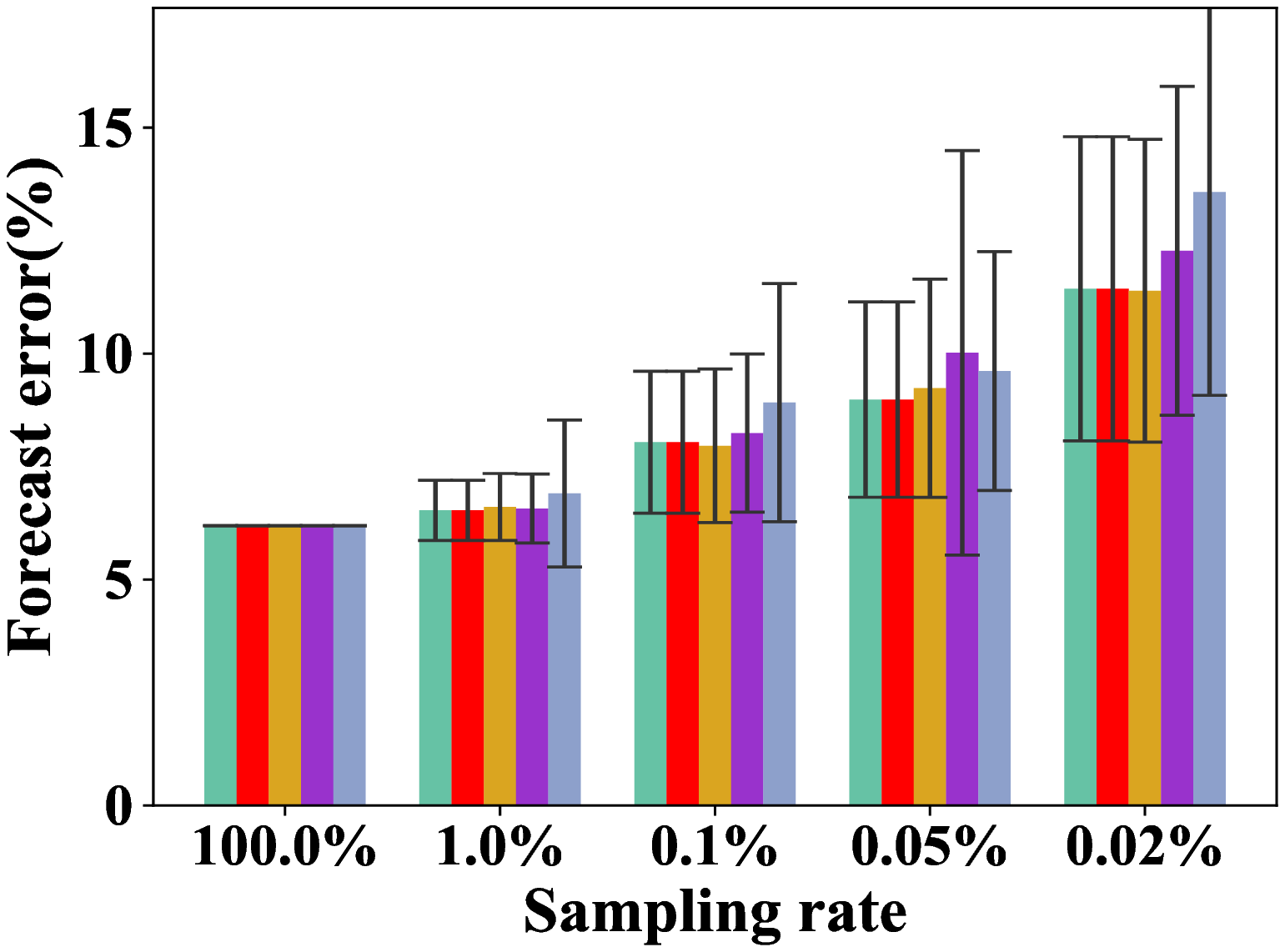}
\end{minipage}
}
\subfigure[LSTM]{\label{fig:exp:ferror:impression:lstm}
\begin{minipage}[t]{0.47\linewidth}
\centering
\includegraphics[width=1.1\linewidth]{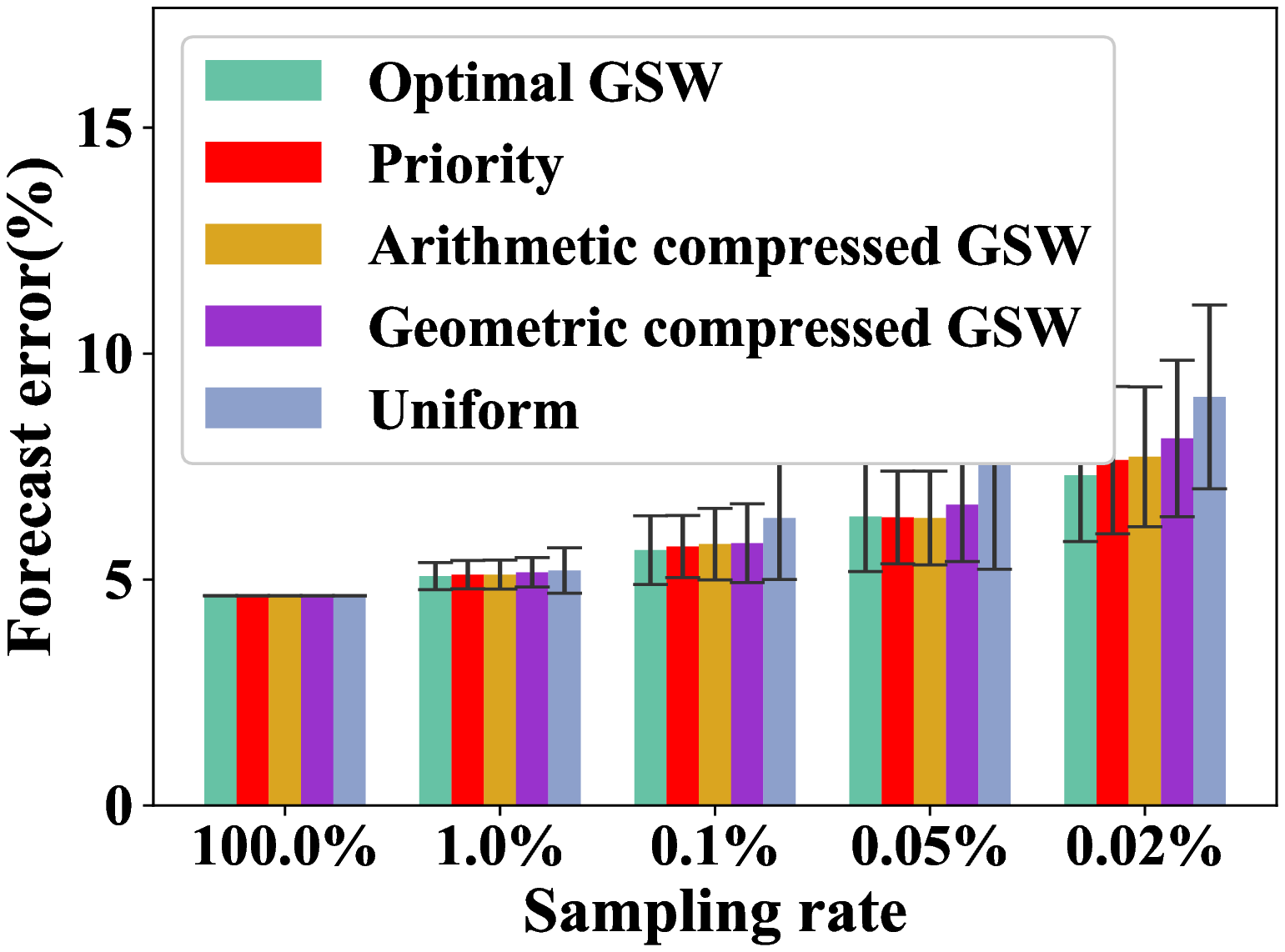}
\end{minipage}
}
\vspace{-0.3cm}
\caption{Forecast error of different sampling methods for varying sampling rate (selectivity 0.5\%, {\sf Impression})}
\vspace{-0.2cm}
\label{fig:exp:ferror:impression}
\end{figure}

\ifdefined\fullversion
	For tasks on $\sf Favorite$ with selectivity $0.5\%$-$5\%$, \cfig\ref{fig:exp:aerror:favorite} reports aggregation errors, and \cfigs\ref{fig:exp:ferror:favorite}-\ref{fig:exp:ferror:favorite:5} report forecast errors when using ARIMA and LSTM as forecasting models; the results on $\sf Impression$ are plotted in \cfigs\ref{fig:exp:ferror:impression}-\ref{fig:exp:ferror:impression:5}.
\else
	For tasks on $\sf Favorite$ with selectivity $0.5\%$-$5\%$, \cfig\ref{fig:exp:aerror:favorite} reports aggregation errors, and \cfig\ref{fig:exp:ferror:favorite} reports forecast errors when using ARIMA and LSTM as forecasting models; the results on $\sf Impression$ are plotted in \cfig\ref{fig:exp:ferror:impression} (those with selectivity $5\%$ can be found in the full version \cite{url:tr}).
\fi
In terms of both aggregation errors and forecasting errors, Priority and Optimal \samplename are very close and better than the others (indeed, at the cost of storing four samples). 
In some cases, Optimal \samplename is even slightly better than Priority (theoretically optimal). This is because, in Priority, if the measure is above some threshold, a row is included in the sample deterministically, which favors the long tail; however, the long tail may or may not satisfy the constraint specified online in the forecasting task.
Uniform is the worst one which is consistent with its analytical error bound \cite{sigmod:Hellerstein97}. Arithmetic/Geometric compressed \samplename needs only one sample, too; they are better than Uniform, and get very close to Priority and Optimal \samplename when the sampling rate is close to $1\%$. For larger selectivity, with more rows satisfying the constraint included in the samples, every sampler gets better.

\begin{figure}[t]
\vspace{-0.2cm}
\subfigure[Varying sampling rate]{
\label{fig:exp:finterval:favorite:05}
\begin{minipage}[t]{0.47\linewidth}
\centering
\includegraphics[width=1.1\linewidth]{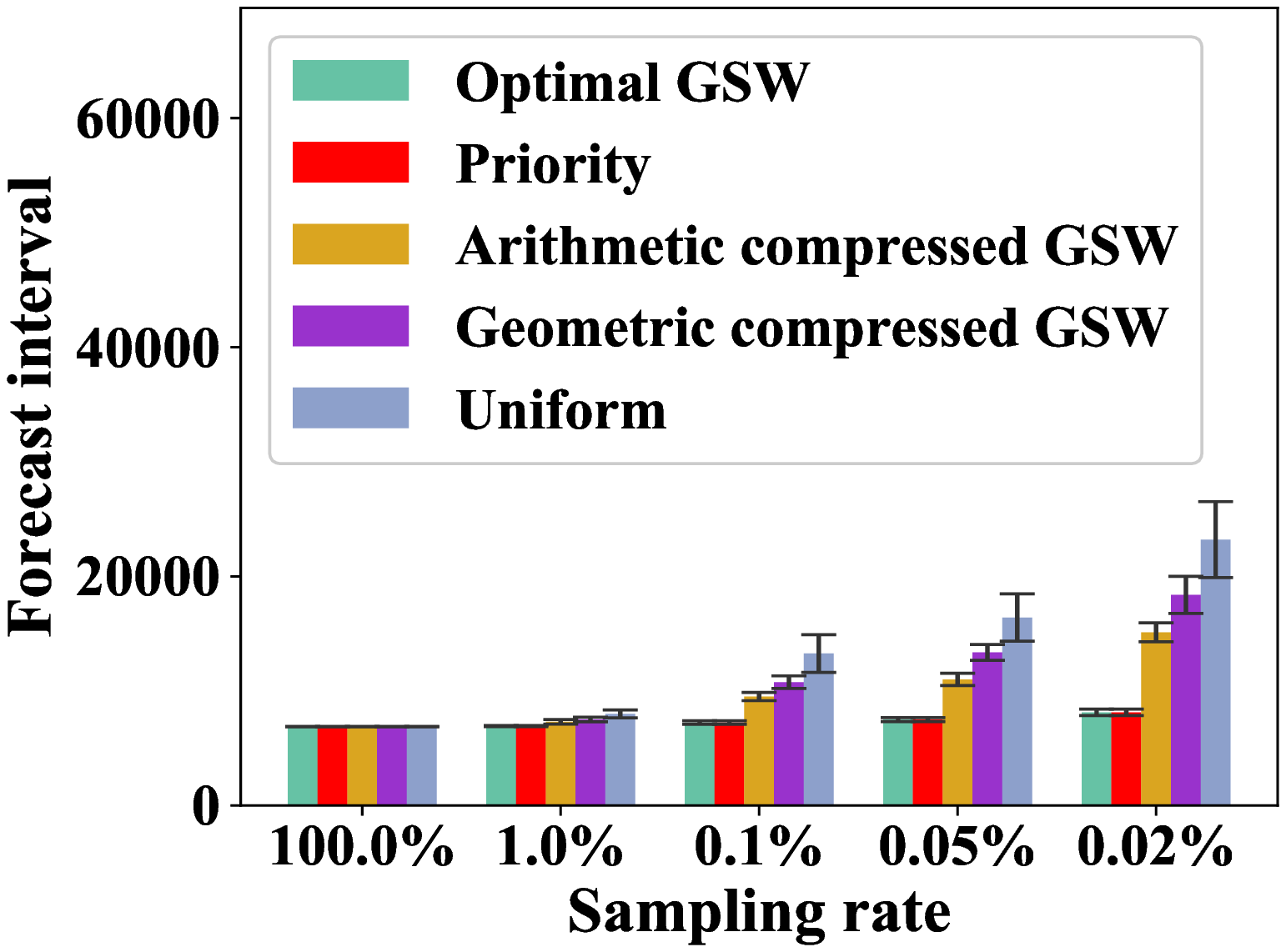}
\end{minipage}
}
\subfigure[Forecast intervals of a real query on 0.02\% sampling]{
\label{fig:exp:finterval:favorite:05:case}
\begin{minipage}[t]{0.47\linewidth}
\centering
\includegraphics[width=1.1\linewidth]{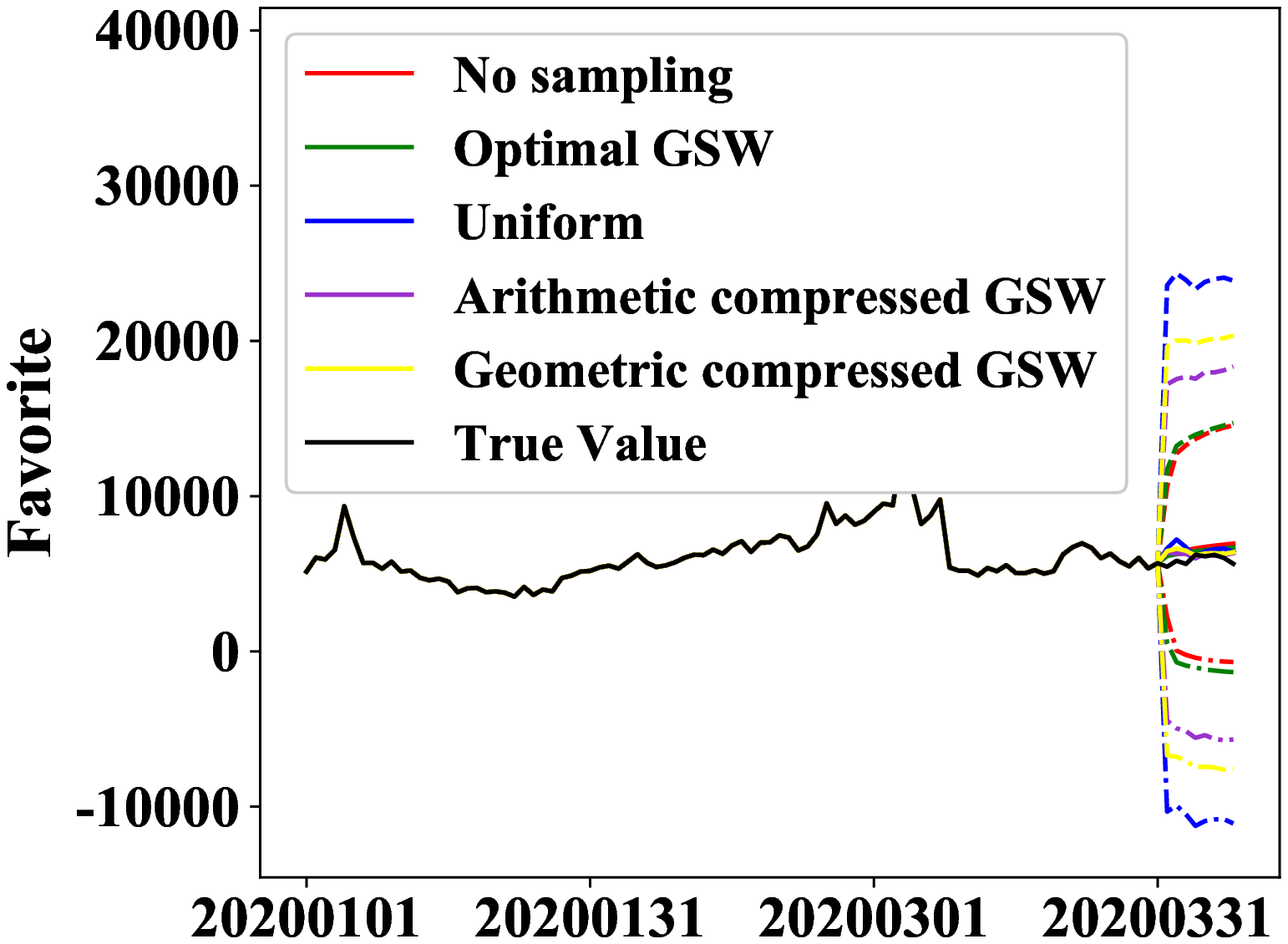}
\end{minipage}
}
\vspace{-0.3cm}
\caption{Forecast intervals (ARIMA) with different sampling methods for varying sampling rate (selectivity 0.5\%, {\sf Favorite})}
\vspace{-0.2cm}
\label{fig:exp:finterval:favorite}
\end{figure}

\cfig\ref{fig:exp:finterval:favorite:05} reports widths of forecast intervals of ARIMA with a confidence level of $90\%$ on measure $\sf Favorite$ for varying sampling rate. With larger sampling rate, every sampler gives narrower forecast intervals, meaning predictions with more confidence. Uniform gives the widest one and Priority and Optimal \samplename give the narrowest ones. \cfig\ref{fig:exp:finterval:favorite:05:case} shows the forecast intervals (dashed lines) for one particular task using different samplers.

In terms of forecasting, LSTM performs consistently better than ARIMA, at the cost of longer response time (as discussed in Exp-II). It is observed that, with increasing sampling rates, when the aggregation error is small enough (\eg, when sampling rate = $1\%$ in \cfigs\ref{fig:exp:aerror:favorite}-\ref{fig:exp:ferror:impression}), it will have little impact on the forecast error/interval in comparison to the case when we use the full data (sampling rate = $100\%$), because it is negligible in comparison to the model and data's noise (\eg, $u_t$ in \eqref{equ:arma} for ARIMA). Forecast errors and forecast intervals have similar trends as aggregation errors. Both observations are consistent with our analytical results in \csec\ref{sec:rtforecast}.


\ifdefined\fullversion
\begin{figure}[t]
\vspace{-0.2cm}
\subfigure[ARIMA]{\label{fig:exp:ferror:impression:5:arima}
\begin{minipage}[t]{0.47\linewidth}
\centering
\includegraphics[width=1.1\linewidth]{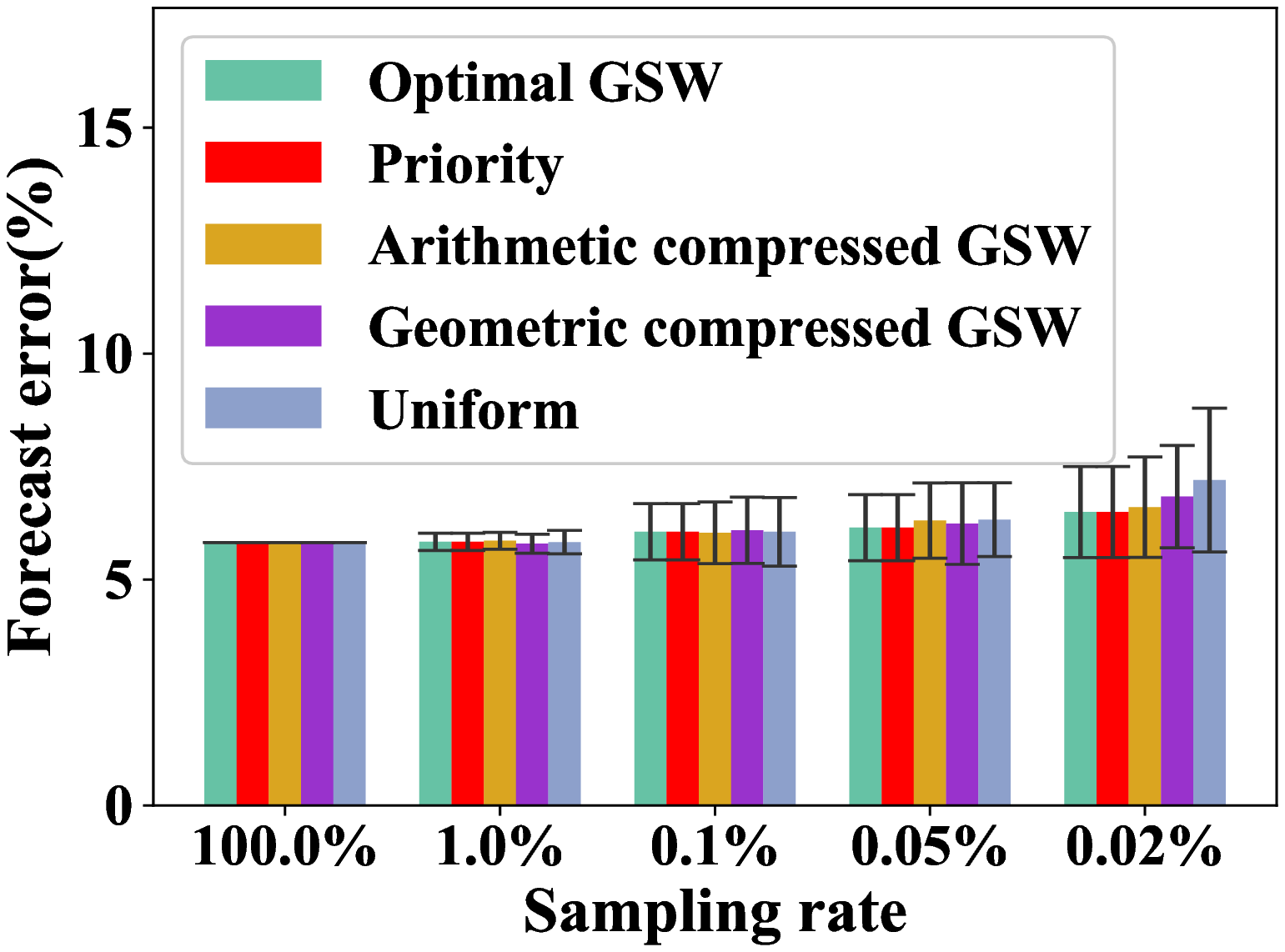}
\end{minipage}
}
%
\subfigure[LSTM]{\label{fig:exp:ferror:impression:5:lstm}
\begin{minipage}[t]{0.47\linewidth}
\centering
\includegraphics[width=1.1\linewidth]{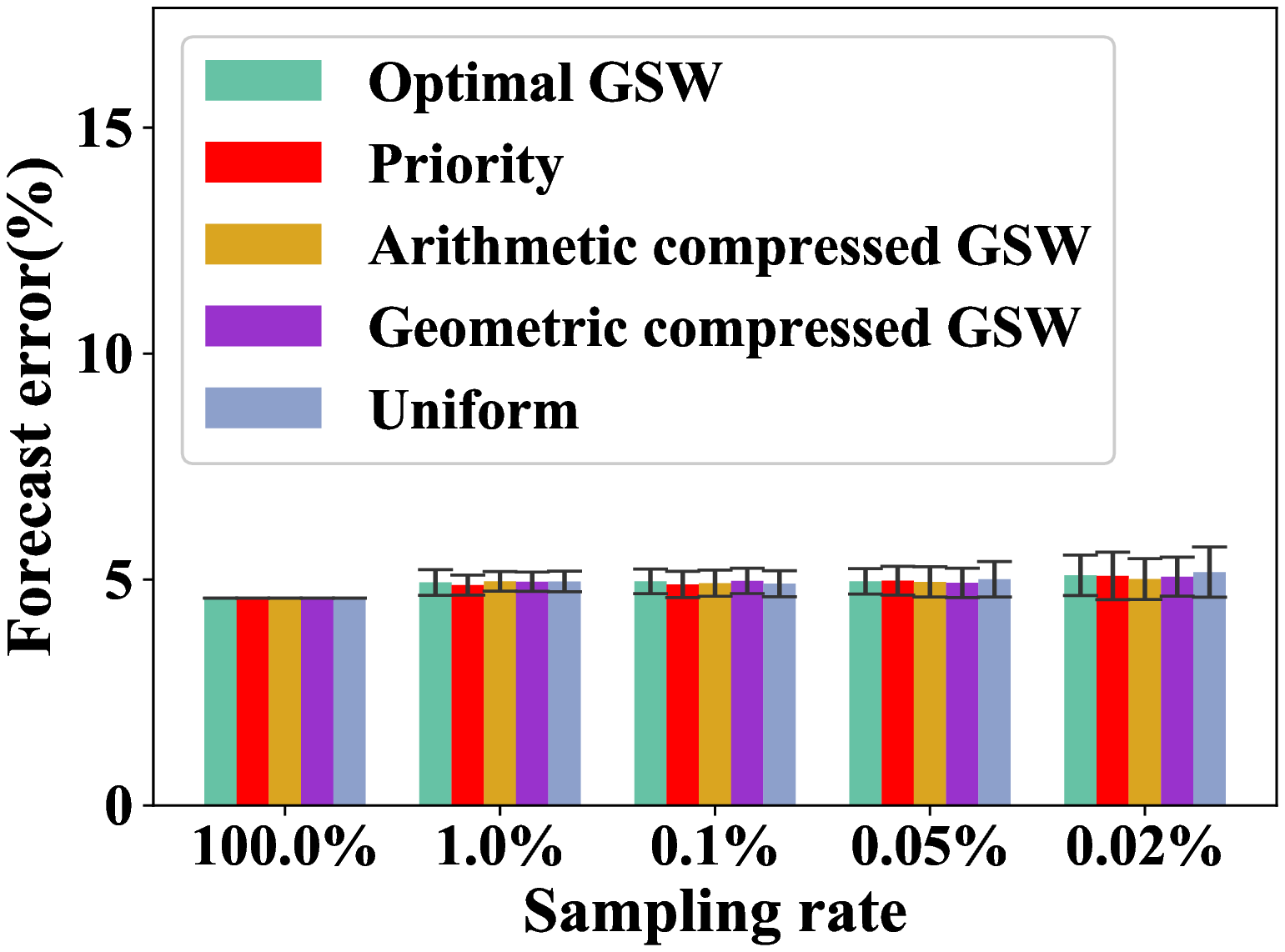}
\end{minipage}
}
\vspace{-0.3cm}
\caption{Forecast error of different sampling methods for varying sampling rate (selectivity 5\%, {\sf Impression})}
\vspace{-0.4cm}
\label{fig:exp:ferror:impression:5}
\end{figure}
\fi

\begin{figure}[t]
\vspace{-0.2cm}
\subfigure[Optimal v.s. Compressed]{
\label{fig:exp:fixacc:samplesize}
\begin{minipage}[t]{0.47\linewidth}
\centering
\includegraphics[width=1.1\linewidth]{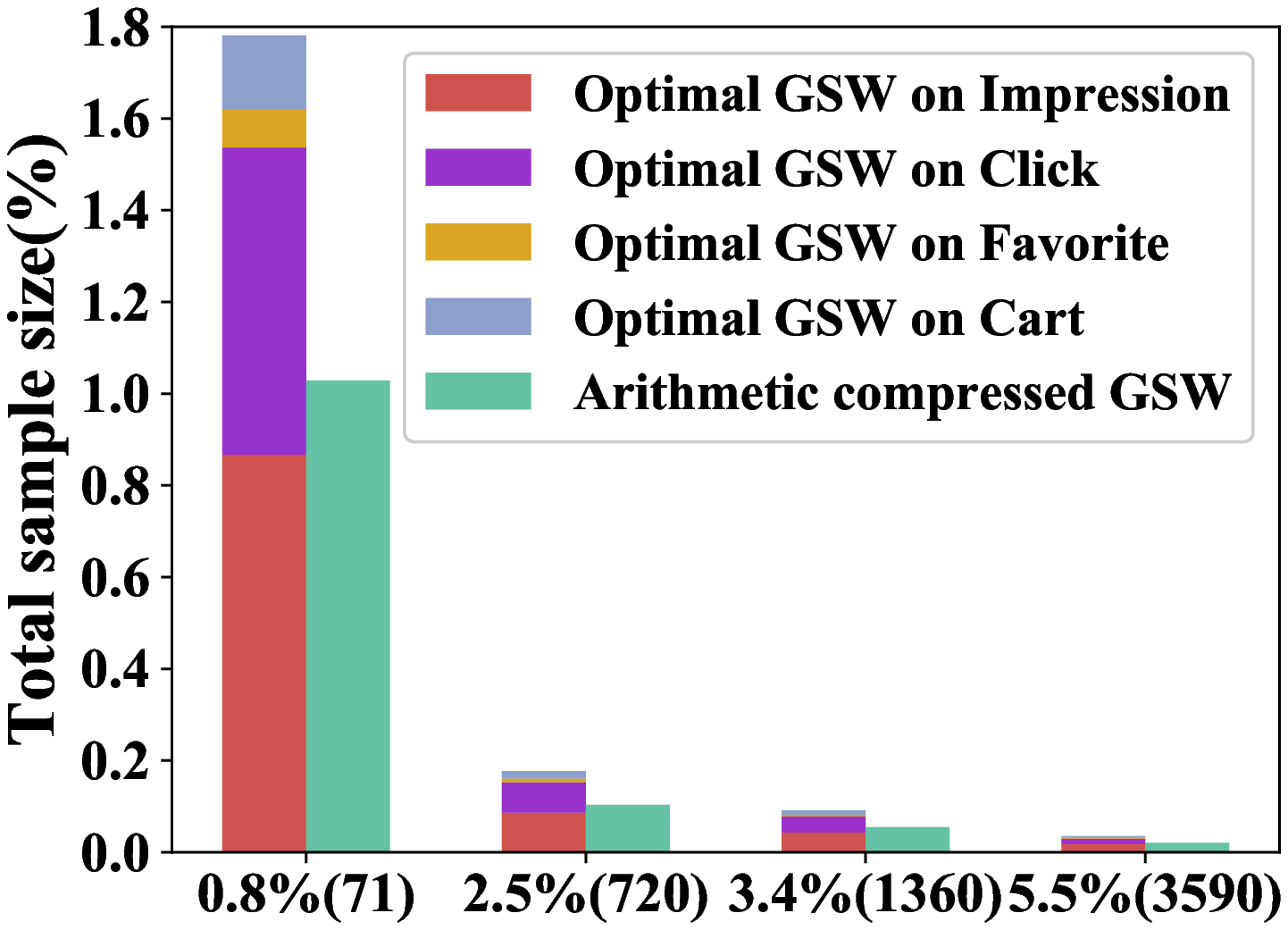}
\end{minipage}
}
\subfigure[Forecast error {(selectivity 5\%)}]{
\label{fig:exp:fixacc:ferror}
\begin{minipage}[t]{0.47\linewidth}
\centering
\includegraphics[width=1.1\linewidth]{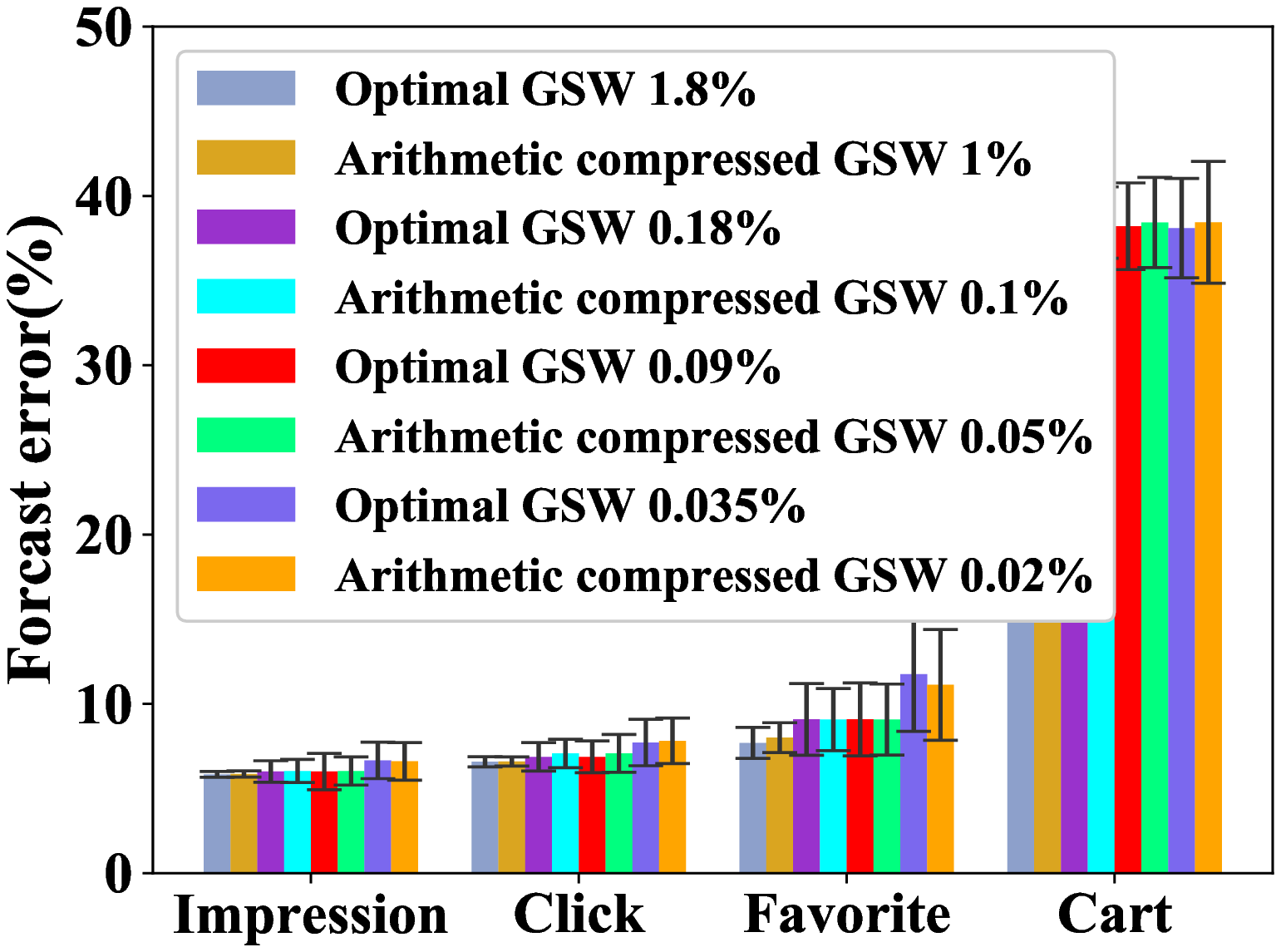}
\end{minipage}
}
\vspace{-0.3cm}
\caption{Space cost for given accuracy requirement}
\vspace{-0.4cm}
\label{fig:exp:fixacc}
\end{figure}

\stitle{Exp-V: Space cost under the same accuracy requirement.}
We evaluate the space cost needed to achieve the same accuracy for different samplers. Since Priority and Optimal \samplename perform similarly, we focus on Optimal \samplename and compare it with Arithmetic compressed \samplename. 
We fix the sample size of Arithmetic compressed \samplename (from $0.02\%$ to $1\%$). And for each measure, we choose the size of an Optimal \samplename sample so that it gives approximately the same aggregation error as Arithmetic compressed \samplename does. In \cfig\ref{fig:exp:fixacc:samplesize}, we report the total size of the four Optimal \samplename samples (the portions for different measures are stacked and labeled with different colors), and the size of the Arithmetic compressed \samplename sample; $x$-axis is the max aggregation error in Arithmetic compressed \samplename with the parameter $\Delta$ in brackets. It shows that, under the same accuracy requirement, the total size of Optimal \samplename samples is around 1.8 times of the size of Arithmetic compressed \samplename.
\cfig\ref{fig:exp:fixacc:ferror} reports forecast errors (of ARIMA) using the samples chosen above on different measures. Due to the way how we choose the sizes of Optimal \samplename samples, Optimal \samplename and Arithmetic compressed \samplename give very close forecast errors.

\section{Related Work}
\label{sec:related}

{\noindent\bf Approximate query processing and other samplers.}
An orthogonal line of work is approximate query processing (AQP), which has been studied extensively during last few decades. One can refer to \cite{sigmod:ChaudhuriDK17} for a comprehensive review. There are two major lines of AQP techniques. i) Online aggregation \cite{sigmod:Hellerstein97,nsdi:CondieCAHES10,pvldb:PansareBJC11} and online sampling-based AQP \cite{sigmod:KandulaSVOGCD16} either assume that the data is randomly ordered, or need to draw random rows from the data table via random I/O accesses which is inefficient in our setting. ii) Offline sampling-based AQP draws offline samples before queries come: some are based on historical workloads \cite{sigmod:Acharya00,VLDB:ganti00,sigmod:Chaudhuri01,TODS:Chaudhuri07,cidr:OlstonBEJR09,cidr:SidirourgosKB11,Eurosys:Agarwal13}, and some are workload-independent \cite{sigmod:Acharya99,sigmod:Acharya00,ICDE:Chaudhuri01,sigmod:BabcockCD03,sigmod:DingHCC016,sigmod:ParkMSW18}. 

The most relevant part in the line of AQP techniques are the samplers proposed to estimate aggregations. We have reviewed such samplers at the beginning of \csec\ref{sec:sampler}.
There are other samplers such as universe (hashed) sampling and stratified sampling introduced in AQP systems \cite{Eurosys:Agarwal13,sigmod:KandulaSVOGCD16,sigmod:ParkMSW18}. These samplers were proposed to handle orthogonal aspects such as missing groups in $\tt Group By$ and joins. They can also be used in our system if we want to extend the task class and data schema we want to support.

\stitle{Precomputing aggregations.}
Another choice is to precompute aggregations or summaries using techniques such as view materialization \cite{VLDB:Agrawal00}, datacube \cite{Springer:Gray1997data}, histograms \cite{PODS:Gilbert01}, and wavelets \cite{sigmod:VitterW99,VLDB:Chakrabarti01, PODS:Gilbert01}. These techniques either have too large space overhead (super-linear) to be applicable for enterprise-scale high-dim datasets, or cannot support complex constraints in our forecasting tasks.

\stitle{Aggregation-forecasting analytics.}
\cite{sigmod:AgarwalCLSV10} studies a very similar problem of forecasting multi-dimensional time series. It considers capturing correlations across the high-dimensional space using either Bayesian models or uniform sampling, and shows that the one based on uniform sampling (which is also evaluated in our experimentes) offers much better forecast accuracy. 
Another relevant line of work is about aggregation-disaggregation techniques \cite{jof:TobiasZ00,book:WestH97} in the forecasting literature. These techniques share some similarity with the Bayesian model in \cite{sigmod:AgarwalCLSV10}, but focus on one-time offline analysis with smaller scale and lower dimensional data.

For multi-dimensional time series, there are works about how to conduct fast similarity search \cite{dasfaa:GongXHCLH15}, but they are less relevant here.

\ifdefined\fullversion
\section{Conclusions}
We introduce a real-time forecasting system \sysname for analyzing time-series relations. There are two core technical contributions, which enable \sysname to handle forecasting tasks on enterprise-scale high-dim time series interactively. First, we analyze how sampling and estimated aggregations on time-series relations enables interactive and accurate predictions. Second, we propose \samplename sampling, a new sampling scheme; we establish a connection from the difference between \samplename's sampling weights and measure values to the accuracy of aggregation estimations. Using this scheme, we can effectively reduce the space (memory) consumption of weighted samples by letting multiple measures share one \samplename sample with the means of these measures' values as sampling weights. 
\else
\subsubsection*{Acknowledgments}
We thank the anonymous reviewers for their helpful comments that improved the quality of the
paper. Zhewei Wei was supported by NSFC No.61832017, No. 61972401 and No. 61932001, by the Fundamental Research Funds for the Central Universities and the Research Funds of Renmin University of China under Grant 18XNLG21, by Beijing Outstanding Young Scientist Program NO. BJJWZYJH01201910002009, and by Alibaba Group through Alibaba Innovative Research Program.
\fi






\ifdefined\fullversion
	\appendix

\section{Missing Details}
\label{app:details}

\subsection{Proof of \cprop\ref{prop:arma11}}
\label{prop:arma11:proof}

We can write the noisy version of ${\rm ARMA(1,1)}$ as
\[
\hat M_t = \alpha_1 \hat M_{t-1}  + (u_t + \beta_1 u_{t-1}) + (\eps_t - \alpha_1 \eps_{t-1}).
\]
Let $\vrinline{u_t} = \sigma^2_u$ and $\vrinline{\eps_t} = \sigma^2_\eps$. We have
\begin{align*}
\epinline{u_t \hat M_t} & = \epinline{u_t^2} + \epinline{u_t(\alpha_1 \hat M_{t-1}  + \beta_1 u_{t-1} + \eps_t - \alpha_1 \eps_{t-1})}
\\
& = \sigma^2_u + 0 = \sigma^2_u,
\end{align*}
from the fact that $u_t$ is dependent only on $\hat M_t$. Similarly,
\[
\epinline{\eps_t \hat M_t} = \sigma_\eps^2.
\]
Taking the variance of both sides of the model, we have
\begin{align*}
\vrinline{\hat M_t} = & \vrinline{u_t + \eps_t} + \vrinline{\alpha_1 \hat M_{t-1} + \beta_1 u_{t-1} - \alpha_1 \eps_{t-1}}
\\
= & \sigma^2_u + \sigma^2_\eps + \alpha_1^2 \vrinline{\hat M_{t-1}} + \beta_1^2 \sigma^2_u + \alpha_1^2 \sigma^2_\eps
\\
& + 2\alpha_1\beta_1\cvinline{\hat M_{t-1}, u_{t-1}} - 2\alpha_1^2\cvinline{\hat M_{t-1}, \eps_{t-1}}
\\
& - 2\beta_1\alpha_1\cvinline{u_{t-1}, \eps_{t-1}}.
\end{align*}
Since $u_{t}$ and $\eps_{t}$ are independent, $\cvinline{\hat M_{t}, u_{t}} = \epinline{u_t\hat M_t}$, and $\cvinline{\hat M_{t}, \eps_{t}} = \epinline{\eps_t \hat M_t}$, we have
\[
\vrinline{\hat M_t} = \alpha_1^2 \vrinline{\hat M_{t-1}}  + (1 + 2\alpha_1\beta_1 + \beta_1^2) \sigma_u^2 + (1-\alpha_1^2) \sigma_\eps^2.
\]
Assuming $\hat M_t$ is weakly stationary ($\vrinline{\hat M_t} = \vrinline{\hat M_{t-1}}$),
\begin{align} \label{equ:arma:var}
\vrinline{\hat M_t} & = \frac{(1+2\alpha_1\beta_1+\beta_1^2) \sigma_u^2 + (1-\alpha_1^2) \sigma_\eps^2}{1 - \alpha_1^2}
\\
& = \frac{1+2\alpha_1\beta_1+\beta_1^2}{1 - \alpha_1^2} \cdot \sigma_u^2 + \sigma_\eps^2 \nonumber
\end{align}
Therefore, the variances of forecasts are linear combinations of $\sigma_u^2$ and $\sigma_\eps^2$, with $\sigma_u^2$ decided by the underlying data distribution, and $\sigma_\eps^2$ decided by sampling scheme and sampling rate. \eop

\subsection{Proof of \cthm\ref{thm:agg}}
\label{thm:agg:proof}
We can first calculate the variance of each ${\hat m_i}$ in \eqref{equ:gbin:variable:probability}:
\[
\vr{\hat m_i} = \frac{m_i^2(\Delta + w_i)^2}{w_i^2}\cdot \frac{w_i}{\Delta + w_i} \cdot \frac{\Delta}{\Delta + w_i} = \frac{\Delta m_i^2}{w_i}.
\]

Since random variables $\hat m_i$'s are independent, we have
\begin{equation} \label{equ:gbin:var}
\vr{\hat{M}} = \sum_{i=1}^n \vr{\hat m_i} = \sum_{i=1}^n\frac{\Delta m_i^2}{w_i}.
\end{equation}

Define $W = \sum_{i=1}^n w_i$. If ${\bf w}$ is $(\lowerb, \upperb)$-consistent with $\meaa$,
\[
\vr{\hat{M}} = \sum_{i=1}^n\frac{\Delta m_i^2}{w_i} \leq \sum_{i=1}^n\frac{\Delta m_i^2}{m_i/\upperb} = \upperb \Delta M.
\]

From \eqref{equ:gbin:variable:probability} and $(\lowerb, \upperb)$-consistency, the expected sample size is
\begin{equation} \label{equ:gbin:samplesize}
\!\ep{|{\mathcal S}_\Delta|} = \sum_{i=1}^n \frac{w_i}{\Delta + w_i} \leq \frac{\sum_{i=1}^n w_i}{\Delta} \leq \frac{\sum_{i=1}^n m_i}{\lowerb \Delta} = \frac{M}{\lowerb\Delta}.
\end{equation}


From the fact that $\hat M$ is unbiased and the above two inequalities,
\begin{align}
\ep{\left(\frac{{\hat M}-M}{M}\right)^2} & \leq \frac{\upperb \Delta M}{M^2} = ({\upperb}/{\lowerb}) \cdot \frac{\lowerb\Delta}{M} \leq \frac{\upperb/\lowerb}{\ep{|{\mathcal S}_{\Delta}|}}. \label{equ:gbin:re}
\end{align}

We can conclude the proof from the above inequality. \eop

\subsection{Finding the Optimal Weight}
\label{app:optweight}

We can choose $\bf w$ to minimize the variance of estimation in \eqref{equ:gbin:var}, while the expected sample size in \eqref{equ:gbin:samplesize} is bounded. That is, solving
\begin{align*}
& \min_{\bf w} \vr{\hat M} = \sum_{i=1}^n\frac{\Delta m_i^2}{w_i}
\\
& \hbox{s.t.} ~ \ep{|S_\Delta|} = \sum_{i=1}^n \frac{w_i}{\Delta + w_i} \leq B.
\end{align*}
Using the method of Lagrange multipliers, we can solve the above problem and obtain the optimal solution $w^*_i$ satisfying:
\[
- \frac{\Delta m_i^2}{{w^*_i}^2} + \frac{\lambda \Delta}{(\Delta + w^*_i)^2} = 0,
\]
where $\lambda$ is the Lagrange multiplier. If we consider a more restrive constraint instead
\[
\sum_{i=1}^n \frac{w_i}{\Delta} \leq B,
\]
we have
\[
- \frac{\Delta m_i^2}{{w^*_i}^2} + \frac{\lambda}{\Delta} = 0,
\]
and thus $w^*_i = \Delta m_i/\sqrt{\lambda}$.

\subsection{Proof of \ccor\ref{cor:agg:geo}}
\label{cor:agg:geo:proof}

%
If we use ${\bf w}^\times$ as the sampling weights for a particular measure $\meaa\supind{p}$, for any row $i \in [n]$, we have
\[
\frac{\meav\supind{p}_i}{w^\times_i} = \frac{\left(\prod_{j=1}^k \meav\supind{p}_{i}\right)^{1/k}}{\left(\prod_{j=1}^k \meav\supind{j}_{i}\right)^{1/k}}= \prod_{j:~j \neq p} \left(\frac{\meav\supind{p}_i}{\meav\supind{j}_i}\right)^{1/k},
\]
and thus,
\[
\max_i \frac{\meav\supind{p}_i}{w^\times_i} = \prod_{j:~j \neq p} \left(\max_i \frac{\meav\supind{p}_i}{\meav\supind{j}_i}\right)^{1/k} = \prod_{j:~j \neq p} \overline\rho^{1/k}_{p,j} \triangleq \upperb;
\]
similarly,
\[
\min_i \frac{\meav\supind{p}_i}{w^\times_i} = \prod_{j:~j \neq p} \underline\rho^{1/k}_{p,j} \triangleq \lowerb.
\]
Therefore, ${\bf w}^\times$ is $(\lowerb,\upperb)$-consistent with $\meaa\supind{p}$, where $\lowerb$ and $\upperb$ are decided by $\underline\rho_{p,q}$'s and $\overline\rho_{p,q}$'s as above, respectively.
From \cthm\ref{thm:agg}, we can derive the following error bound: 
\[
{\rm RE}({\hat M}\supind{p}) \! \leq \! {\rm RSTD}({\hat M}\supind{p}) \! \leq \sqrt{\frac{\prod_{j: ~ j \neq p}\rho_{p,j}^{1/k}}{\ep{|{\mathcal S}_{\Delta}|}}} \leq \sqrt{\frac{(\rho^{1/k})^{k-1}}{\ep{|{\mathcal S}_{\Delta}|}}}
\]
which concludes the proof. \eop

\subsection{Proof of \cprop\ref{prop:consistencyl2}}
\label{prop:consistencyl2:proof}

Since ${\bf w}$ $=$ $[w_i]$ are $(\lowerb,\upperb)$-consistent with measure $\meaa = [\meav_i]$,
\[
\meav'_i = \frac{\meav_i}{\sum_j \meav_j} \leq \frac{\upperb w_i}{\sum_j \lowerb w_j} = \frac{\upperb}{\lowerb} \cdot w'_i = \theta \cdot w'_i,
\]
where $\theta = \upperb/\lowerb$. Similarly,
\[
\meav'_i \geq \frac{1}{\theta} \cdot w'_i.
\]
Putting them together, we have
\[
|\meav'_i - w'_i| \leq \max\{(\theta - 1), (1 - 1/\theta)\} \cdot w'_i.
\]
Since $\theta \geq 1$, we have $\theta - 1 \geq 1 - 1/\theta$, and thus
\[
|\meav'_i - w'_i| \leq (\theta - 1) \cdot w'_i.
\]
Summing them up,
\[
\|\meaa' - {\bf w}'\|_1 = \sum_i |\meav'_i - w'_i| \leq (\theta - 1) \cdot \sum_i w'_i = (\theta - 1).
\]
\eop
\else
\fi


\end{document}